\documentclass[5p,times]{elsarticle}

\usepackage{ecrc}
\volume{00}
\firstpage{1}
\journalname{Astronomy and Computing}
\runauth{}
\jid{procs}
\usepackage{amssymb}
\usepackage[figuresright]{rotating}
\usepackage{tabularx}
\usepackage{booktabs}
\usepackage{graphicx}
\usepackage{blindtext}
\usepackage{multirow}
\usepackage{amsmath}
\usepackage[singlelinecheck=false]{caption}
\usepackage[ruled,vlined]{algorithm2e}

\makeatletter
\def\@author#1{\g@addto@macro\elsauthors{\normalsize%
    \def\baselinestretch{1}%
    \upshape\authorsep#1\unskip\textsuperscript{%
      \ifx\@fnmark\@empty\else\unskip\sep\@fnmark\let\sep=,\fi
      \ifx\@corref\@empty\else\unskip\sep\@corref\let\sep=,\fi
      }%
    \def\authorsep{\unskip,\space}%
    \global\let\@fnmark\@empty
    \global\let\@corref\@empty  
    \global\let\sep\@empty}%
    \@eadauthor={#1}
}
\makeatother

\begin{document}

\begin{frontmatter}

\dochead{}

\title{Data Compression in the Petascale Astronomy Era: \\a GERLUMPH case study}

\author{Dany Vohl\corref{cor1}}
\ead{dvohl@astro.swin.edu.au}
\cortext[cor1]{Corresponding author}

\author{Christopher J. Fluke}
\ead{cfluke@astro.edu.au}

\author{Georgios Vernardos}
\ead{gvernardos@astro.swin.edu.au}

\address{
Centre for Astrophysics \& Supercomputing,\\
Swinburne University of Technology, PO Box 218, \\
Hawthorn, Victoria, 3122, Australia}

\begin{abstract}
As the volume of data grows, astronomers are increasingly faced with choices on what data to keep --- and what to throw away. Recent work evaluating the JPEG2000 (ISO/IEC 15444) standards as a future data format standard in astronomy has shown promising results on observational data. However, there is still a need to evaluate its potential on other type of astronomical data, such as from numerical simulations. GERLUMPH (the GPU-Enabled High Resolution cosmological MicroLensing parameter survey) represents an example of a data intensive project in theoretical astrophysics. In the next phase of processing, the $\approx27$ terabyte GERLUMPH dataset is set to grow by a factor of 100 --- well beyond the current storage capabilities of the supercomputing facility on which it resides. In order to minimise bandwidth usage, file transfer time, and storage space, this work evaluates several data compression techniques. Specifically, we investigate off-the-shelf and custom lossless compression algorithms as well as the lossy JPEG2000 compression format. Results of lossless compression algorithms on GERLUMPH data products show small compression ratios (1.35:1 to 4.69:1 of input file size) varying with the nature of the input data. Our results suggest that JPEG2000 could be suitable for other numerical datasets stored as gridded data or volumetric data. When approaching lossy data compression, one should keep in mind the intended purposes of the data to be compressed, and evaluate the effect of the loss on future analysis. In our case study, lossy compression and a high compression ratio do not significantly compromise the intended use of the data for constraining quasar source profiles from cosmological microlensing. 
\end{abstract}


\end{frontmatter}

\section{Introduction}

\subsection{The Petascale Astronomy Era}
The expression \emph{Big Data} is currently a popular one in astronomy. It is also broadly and liberally used. Recognising that ``bigness'' is an arbitrary concept related to the usual data sizes and volumes within a sub-field of astronomy, we can consider three types of Big Data. The first case happens for projects dealing with \emph{lots of small to medium sized files}. This is typical of many modern observational survey programs \citep[e.g.][]{Ahn2012}. Each image is only a few hundred megabytes (MB) in size, but many thousands of individual images are recorded and archived. The second case is when there are a \emph{few very big files}. Here belong cases such as the Millennium simulation \citep{Springel-2005}. The full particle data was stored at 64 time steps, each of size 300 gigabytes (GB), giving a raw data volume of nearly 20 terabytes (TB). Finally, the third case is when there are \emph{lots of very big files}. For example, this is the situation that will occur with the large spectral cubes from instruments such as MeerKAT \citep{Booth2009} and ASKAP \citep{Guzman2010}. Similarly, the Square Kilometer Array is expected to gather 14 exabytes of data and store about one petabyte (PB\footnote{1 petabyte = $10^{15}$ bytes.}) every day \citep{lazio-2011, ibm-2013, Quinn2015}.

With this growth of data volume in science come new challenges. How to efficiently store such data in data storage facilities? How to transmit such data over a network in an acceptable time interval through limited bandwidth? How to perform visualisation and analysis on large samples of files?

Visualisation and analysis of terabyte-scale data is already a challenge for existing astronomical softwares and the current work practices of astronomers \citep{HassanFluke2011}. Furthermore, such large data volumes cannot be processed, stored or viewed as a whole on desktop computers, even taking into account projected advances for hard drives and network technologies \citep{kitaeff-2012}. Research in a variety of sub-disciplines of astronomy is underway to solve such issues \citep[e.g.][]{Anderson2011, Broekema2012, Hassan-2012, Hassan-2013}. 

As the volume of data grows, astronomers are increasingly faced with choices on what data to keep --- and what to throw away. In this work, we concentrate on the questions regarding storage and transmission by investigating \emph{data compression}.

\subsection{Data Compression}
\label{subsection::datacompression}

Data Compression emerged from the work of \citet{shannon} on Information Theory. One of the critical ideas of Shannon's theory was the relation between the probability of occurrence of a particular value $v$ in a data source $S$ of length $n$ and the amount of information $i$ that this value carried. If $v$ occurs $f(v)$ times, the probability $p(v)$ of observing $v$ in $S$ is given by 

\begin{equation}
p(v) = \frac{f(v)}{n}. 
\end{equation}

The amount of information associated with $v$ is given by 
\begin{equation}
i(v) = \log \frac{1}{p(v)} = -\log p(v). 
\end{equation}

Averaging over all the information included in $S$ such that 

\begin{equation}
H(S) = -n \sum_{v=1}^{n} p(v) \log p(v) 
\end{equation}

is the entropy of the source (often called Shannon entropy), a concept closely linked to Ludwig Boltzmann's second law of thermodynamics which stipulates that the entropy of a system can only increase \citep{frank}. It is a lower bound on the space required to represent a source of information perfectly \citep{shannon}. 


Compression is achieved by using an encoder to remodel the original data in a compressed manner, bringing it close to or even beyond its entropy. Later, a decoder is used to uncompress the compressed data back to its original form. We can note that the concept of entropy naturally divides data compression into two main paradigms: \emph{lossless compression}, where the output data of the decoder is identical to the encoder's input data, and \emph{lossy compression}, where the output of the decoder is different to the encoder's input \citep{DCC}  --- it is an approximation of the input. 

\subsection{Big Data, data format, and data compression}
\label{subsection::datacompression}

The idea of minimizing the volume of astronomical data using data compression can be traced back to the 1970s \cite[e.g.][]{Labrum1975, Miller1976} and such techniques have been used and developed ever since \cite[e.g.][]{Landau1984, Andrianov1984, Cafforio1985, Press1992, white-1994, Veran1994, Starck1995, Vasilyev1998, Gaudet2000, Pence2000, Fixsen2000, Pence2011, Seaman2011}. 

In the context of the petascale astronomy era, prior work considered questions such as: which data format/model should be used to enable work with Big Data (e.g. \citealp{kitaeff-2012}; \citealp{Kitaeff-2014}, this issue; \citealp{Natusch2014}; \citealp{Mink2014}; \citealp{Price2014}); what bandwidth to keep in case of lossy compression \citep{Price-Whelan-2010}; and whether lossy compression affects analysis \citep[e.g.][]{white-1994, Lior2005, Pence2010, Vohl:Thesis:2013, Peters-2014}. 

Recent work by \citet{kitaeff-2012}, \citet[this issue]{Kitaeff-2014}, and \citet{Peters-2014} investigates whether the JPEG2000 (ISO/IEC 15444) standards could be adopted more generally within astronomy. Within the JPEG2000 specification are features attractive to astrophysics such as: progressive transmission; the ability to decode only part of the data without having to load it all into memory (useful when dealing with larger-than-memory files); and the possibility to include customized metadata.

In the context of radio astronomy, \citet{Peters-2014} and \citet[this issue]{Kitaeff-2014} showed that the data format prescribed by the JPEG2000 standard can deliver high compression ratios with limited error on analysis of observational data\footnote{\citet{Peters-2014} actually used synthetic spectral-imaging data-cubes for their experiment allowing them to have a complete knowledge about the sources in the cube.}. \citet{Peters-2014} evaluated the effect of the lossy compression on analysis. To do so, they ran a source finder on a spectral-imaging data-cube pre and post compression and compared the soundness of the outcome. They showed that using the JPEG2000 format, the strongest sources ($\gtrsim2000$ mJy km/s and higher) may still be retrievable at extremely high compression ratio; in such cases, the compressed file would be more than $15,000$ times smaller than the original file. In cases where they used a high quantization step during the compression (giving them compression ratio of about $\lesssim90$:$1$), JPEG2000 enabled them to identify low integrated flux sources (less than 800 mJy km/s). They concluded that the compression is denoising the cube, allowing sources previously obscured by noise to be identified.  

These results are promising for the observational community. However, to our knowledge, there is not yet any systematic work on other types of astrophysical data using the JPEG2000 standard, such as data from numerical simulations. Observational data is noisy by nature (e.g. the instrument capturing the data is imperfect). However, for numerical data, one would expect that lossy data compression will tend to introduce additional noise. It may therefore be considered an inherently negative aspect of such format for theorists who do not want to corrupt the data that has been carefully generated. But what actually is the effect of such compression on theoretical data? To investigate this question, we selected the Graphics Processing Unit-Enabled High Resolution cosmological MicroLensing parameter survey (GERLUMPH) theoretical dataset \citep{VerdanosEtAl2014} as a case study. 

\subsection{Case study: GERLUMPH}
\label{subsection::gerlumph}

GERLUMPH represents a recent example of a data intensive project in theoretical astrophysics. It has been designed to study \emph{quasar microlensing}: the gravitational lensing effect of stellar mass objects within foreground galaxies that lie along the line of sight to multiply-imaged background quasars. There are currently $\approx90$ known multiply imaged quasars, but an increase to a few thousands is anticipated to happen soon due to the commencement of synoptic all-sky surveys \citep{Oguri2010}. GERLUMPH has been developed to explore and understand the quasar microlensing parameter space (defined in terms of the external shear, $\gamma$, and the convergence, $\kappa$) by providing a theoretical resource in preparation for these future discoveries.

Quasar microlensing is usually studied numerically using magnification maps: pixelated versions of the caustic pattern in the background source plane created by the foreground microlenses.

GERLUMPH accelerated the process of generating cosmological magnification maps by using graphic processing units \citep[GPU; see][]{Thompson-2010} resulting in the generation of $\approx$ 70,000 maps in a short period of time \citep{VerdanosEtAl2014}. This work was completed on the GPU Supercomputer for Theoretical Astrophysics Research (gSTAR) national facility hosted at Swinburne University of Technology. Each map is composed of $10,000^2$-pixel requiring $\approx400$ MB per map. This corresponds to a manageable $\approx27$ TB of data for the parameter survey.

Observational properties of microlensed quasars can be obtained by performing a convolution between a quasar model and a magnification map. The next stage of GERLUMPH processing will involve convolution of 100 different physically-motivated quasar profiles with all of the GERLUMPH maps. It will produce a dataset two orders of magnitude larger, which would exceed the current available GERLUMPH storage space on gSTAR (currently $\approx3.4$ PB for all projects). While it is desirable to save the convolved maps for future use --- can they actually be stored?

\subsection{Magnification maps}
\label{section::magnification_maps}

To understand the opportunities for compression, we need to understand the properties of the data. GERLUMPH generates the magnification maps using a technique called Inverse Ray-Shooting \citep{Kayser1986, Schneider1986, Schneider1987, Thompson-2010}. The technique proceeds by shooting a large number of light rays ($\approx 10^9$) from the observer through the lens plane, where they are deflected by $N_*$ individual microlenses (compact, point-mass objects), defined as 

\begin{equation}
N_* = \frac{\kappa_*A}{\pi \langle M \rangle}.
\label{eq::Nstar}
\end{equation} 

Here $\kappa_*$ is the convergence caused by compact objects, $\langle M \rangle$ is the mean mass of the microlenses, and A is the area where they are distributed. The total convergence in cosmological microlensing has contributions from compact objects, $\kappa_*$, and smooth matter, $\kappa_s$, such that:

\begin{equation}
\kappa=\kappa_s+\kappa_*.
\label{eq:convergence}
\end{equation}

%
%

The gravitational lens equation, used in the context of cosmological microlensing by $N_*$ microlenses of mass $m$ in the presence of a smooth matter distribution and an external shear, $\gamma$, is defined as:

\begin{equation}
\boldsymbol{y} = \left( \begin{array}{cc}
1-\gamma & 0  \\
0 & 1-\gamma \end{array} \right) \boldsymbol{x} - \kappa_s\boldsymbol{x} - \sum_{i=1}^{N_*}m_i\frac{(\boldsymbol{x}-\boldsymbol{x_i})}{|\boldsymbol{x}-\boldsymbol{x_i}|^2}.
\label{eq::grav_lens}
\end{equation}

The quantities $\textbf{x}$ and $\textbf{y}$ are two-dimensional vectors in the lens and source plane respectively, $x_i$ are the location of the $N_*$ individual lenses with mass $m_i$. Each projected light ray from the lens plane to the source place is accumulated on a grid: the magnification map. The characteristic microlensing scale length in the source plane is the Einstein radius defined as:

\begin{equation}
R_{Ein} = \sqrt{\frac{D_{os}D_{ls}}{D_{ol}} \frac{4G\langle M \rangle}{c^2}}
\end{equation}

where $D_{ol}$, $D_{os}$, and $D_{ls}$, are the angular diameter distances from observer to lens, observer to source, and lens to source respectively, $\langle M \rangle$ is the mean mass of pointÐmass microlenses, $G$ is the gravitational constant, and $c$ is the speed of light. A typical range of values for $R_{Ein}$ can be obtained from the sample of 87 lensed quasars compiled by \citet{Mosquera2011}: $5.11 \pm 1.88 \times 10^{16}$cm. In this paper, we use the mean from the \citet{Mosquera2011} sample as a typical value for $R_{Ein}$. Therefore, the pixel size of the high-resolution GERLUMPH maps corresponds to $\approx 1.28 \times 10^{14}$ cm.

The number of light rays that reach a pixel of the source plane $N_{i,j}$ at coordinate $i,j$, compared to the number of rays that would reach each pixel if no microlensing was occurring, $N_{avg}$, gives the local magnification value, $\mu_{i,j}$:

\begin{equation}
\mu_{i,j} = \frac{N_{i,j}}{N_{avg}}.
\label{eq::local_magnification_value}
\end{equation} 

\begin{figure*}[!htb]
\setlength{\tabcolsep}{0.07cm}
\begin{tabular}{cc}
\includegraphics[width=9.053cm]{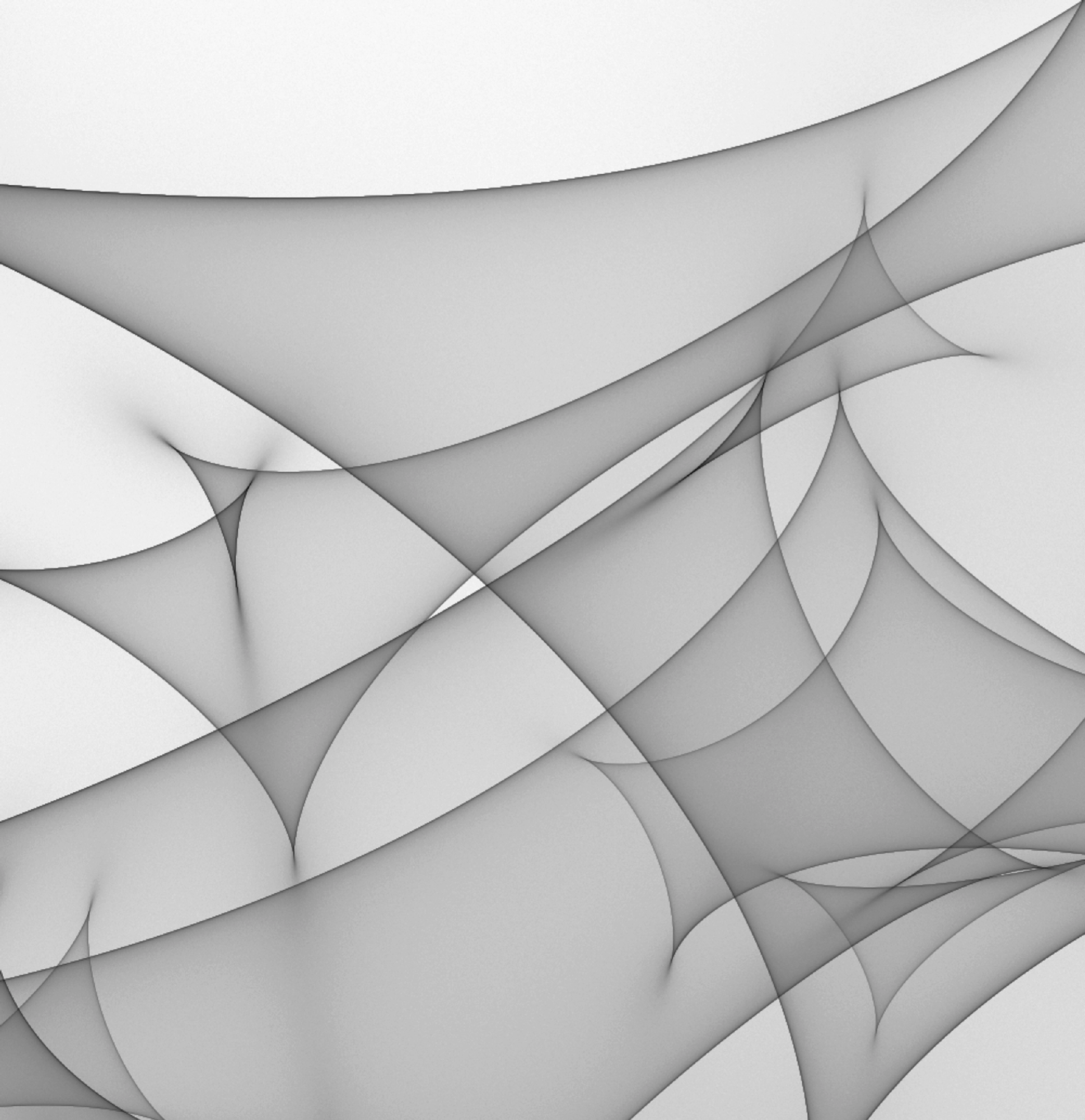} &
\includegraphics[width=9.053cm]{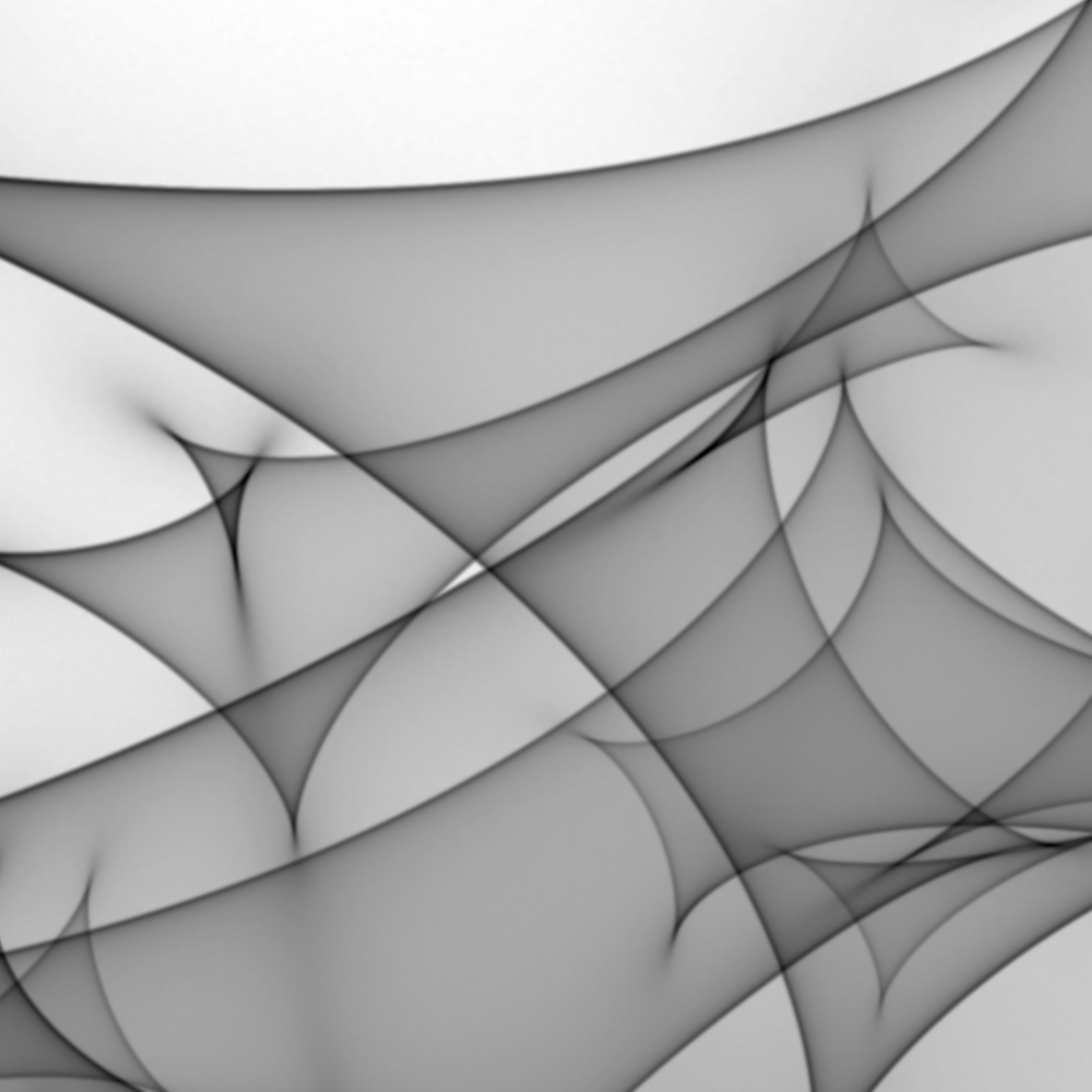} \\
\includegraphics[width=9.053cm]{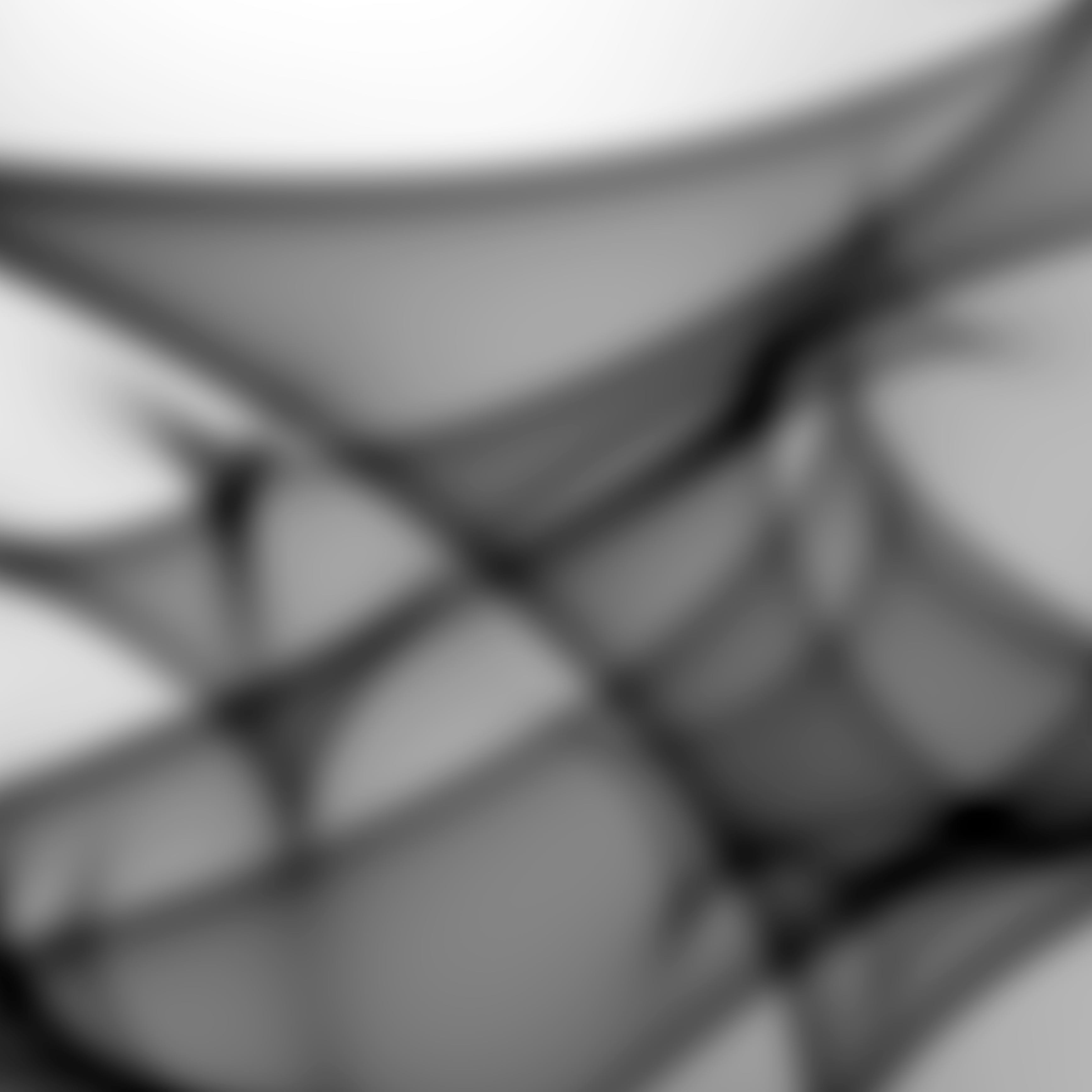} &
\includegraphics[width=9.053cm]{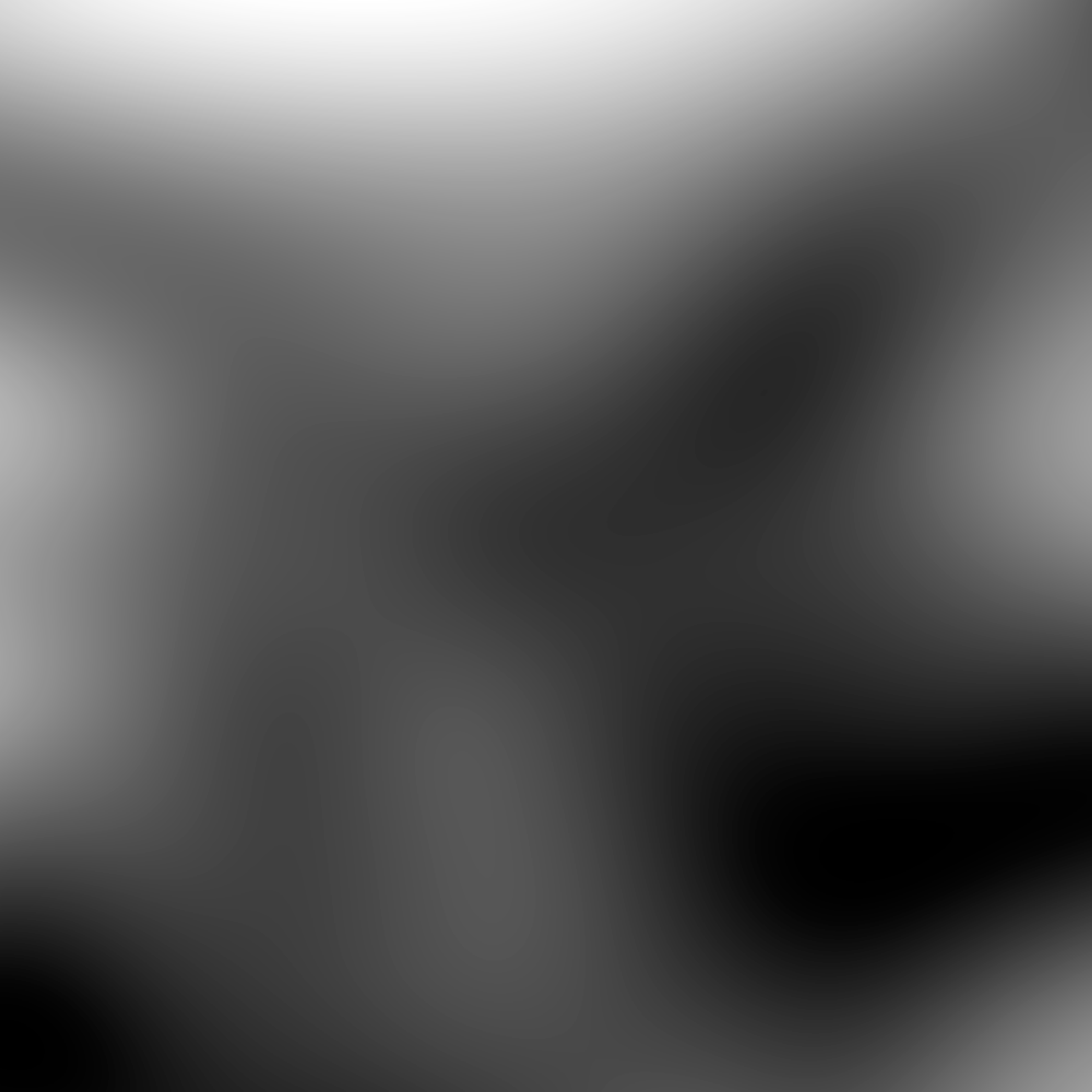}
\end{tabular}
\caption{Examples of GERLUMPH magnification maps. Each image is a close-up portion of $1000^2$-pixel, equivalent to 2.5 $R_{Ein}$. The upper left map is an original map. Upper right, lower left and lower right quadrant show the original map convolved with a $10^2$-pixel, $100^2$-pixel, and $500^2$-pixel quasar profiles respectively. The quasar profile widths correspond to 0.1\%, 1\% and 5\% of the map width.}
\label{fig::maps}
\end{figure*}

In this work, we use two types of maps. A point-source magnification map (hereafter original map) is represented as an array of integers where each point is the ray count ($N_{i,j}$) at a given pixel in the map. A magnification map convolved with a quasar profile (hereafter convolved map) is a matrix of floats representing a magnification factor ($\mu_{i,j}$) at a particular pixel. An example of such maps is shown in Figure \ref{fig::maps}. 

Convolved maps are slightly smaller than original maps. To avoid introducing distortion from the periodic nature of the convolution when applied to our non-periodic maps, a frame is removed from the map after the convolution process. The width of each cropped section is half of the quasar profile width, i.e. a $10,000^2$-pixel original map convolved with a $500^2$-pixel quasar profile will become a $9500^2$-pixel map.



\newcommand{\figurescaletwo}{0.256}


\section{Lossless data compression experiment}
\label{section::losslessExperiment}
Starting with the idea that data corruption is an undesirable outcome, we first evaluate lossless compression. We compare results from popular lossless compression algorithms with lossy compression. We evaluated four lossless compression software commonly found on most linux distributions, as well as two other lossless compression solutions ($\mathtt{LZ4}$ and $\mathtt{bitpacking}$) in order to evaluate how well such techniques compress the GERLUMPH data. 


\subsection{Evaluated softwares}
$\mathtt{Gzip}$ (version 1.3.12) is part of the family of compression techniques called ``dictionary methods'', in which a dictionary of substrings is constructed to represent the data. $\mathtt{Gzip}$ implements the Deflate algorithm\footnote{http://tools.ietf.org/html/rfc1951}, a combination of the classic LZ77 algorithm \citep{LZ77} and static Huffman coding \citep{huffman}. The main idea behind LZ77 is to reduce the redundancy of the data by replacing values, or preferably a sequence of values, with metadata referencing to a dictionary of recurrent sequences of values seen previously by the encoder. Encoding sequences of values as blocks enables the encoder to take advantage of mutual information in neighbouring values. Huffman coding is an entropy coder based on uniquely decodable, variable length, prefix codes that associates shorter codes for frequent values and longer codes for rare ones. 

$\mathtt{bzip2}$ (version 1.0.5) is a mixture of three techniques \citep{DCC}. The first technique is the Burrows--Wheeler transform \citep{burrows-1994}, a reversible transform used as pre-processing for compression. It sorts blocks of data, simplifying the structure of input data so as to enable more effective compression. The second is the move-to-front \citep{Bentley1986} technique, a locally adaptative method, adapting to the frequencies of symbols in local areas of the data. Finally, the last technique is the Huffman coding described above.

$\mathtt{LZMA}$ (version 4.999.9beta) stands for Lempel--Ziv--Markov chain-Algorithm. It is also a LZ77 variant based on a large search buffer, a hash function that generates indexes, and 2 search methods: the fast method, which uses a hash-array of lists of indexes and the normal method, which uses a hash-array of decision trees \citep{DCC}. $\mathtt{xz}$ (version 4.999.9beta) is a multi-functionality software that incorporates the $\mathtt{LZMA}$ algorithm. It not only enables compression of a single file, but also enables archiving functionality (many files into one compressed file). We evaluated this software to compare if the implementation differs from the standalone $\mathtt{LZMA}$. 

$\mathtt{LZ4}$ (version r101) is a recent popular and really fast variant of $\mathtt{LZ77}$, designed for parallel computing. The design principle behind this software is simplicity delivered through a simple code that delivers a fast execution \citep{lz4}. It generally delivers lower compression ratios than the previously mentioned techniques, but is rather meant to have a much smaller runtime, which can be of interest when one needs to compress a great number of files. 

The $\mathtt{bitpacking}$ implementation\footnote{We modified the fast C++ implementation of Daniel Lemire (http://pastebin.com/ugGnk00p) to work with our data. More information can be found at http://lemire.me/blog/archives/2012/03/06/how-fast-is-bitpacking/} was evaluated following a discussion in \citet{Vernardos-2014} stating that disk space could be saved using less than 32 bits to encode integers, as not all maps require 32-bit to represent the necessary raw information. Instead of defining our own integer types and therefore dealing with the computation overhead, we investigated how much space can be saved by ``bit-packing'' into 32-bit buffers the magnification maps for which less than 32-bit is required, as described in \citet{Lemire-2012}. The software evaluates the number of bits required to represent all values in a magnification map as a function of $\mathtt{b = \lceil log_2(max) \rceil}$, where $\mathtt{max}$ is the maximum value in the file. If $\mathtt{b < 32}$, it packs $\mathtt{n \times b}$ into $\mathtt{\lceil \frac{n \times b}{32} \rceil}$ 32-bit buffers, where $\mathtt{n}$ is the number of values to encode.\\

\subsection{Methodology}
\label{section::methodology_lossless}

For each lossless technique mentioned in the previous section, we obtained the compression ratio (\#:1) where 

\begin{equation}
\# = \frac{size_{original}}{size_{compressed}},
\end{equation}

\noindent and the median time in seconds (s) of three rounds of compression and decompression. We selected 306 maps uniformly distributed over the $\kappa,\gamma$ parameter space covered by GERLUMPH-GD1. We used two different types of maps: original maps and maps convolved with 3 different quasar profile widths: $10$-pixel, $100$-pixel, and $500$-pixel corresponding to 0.1\%, 1\% and 5\% of the map width respectively (Figure \ref{fig::maps}). This resulted in $306\times4=1224$ maps per compression technique. 

The different data types (integer, float) have an effect on general purpose lossless compression algorithms. By evaluating compression over parameter space, it enables us to evaluate variations in compression ratio and runtime relative to map types. 



All techniques tested in this section, except $\texttt{bitpacking}$, includes flags which can be set to obtain different compression range. The flags let the user choose between lower to higher compression ratio. Lower compression ratio is faster at runtime while higher compression takes longer to run. We tested these extreme values for all techniques. All lossless experiments were conducted using Linux (CentOS release 6.6) on SGI C2110G-RP5 nodes containing 2 eight-core SandyBridge processors at 2.2 GHz, where each processor is 64-bit 95W Intel Xeon E5-2660, with 10 GB of RAM allocated.

\subsection{Results}
\label{section::results_lossless}

\begin{figure*}[!htb]
\centering
\includegraphics[width=18.4cm]{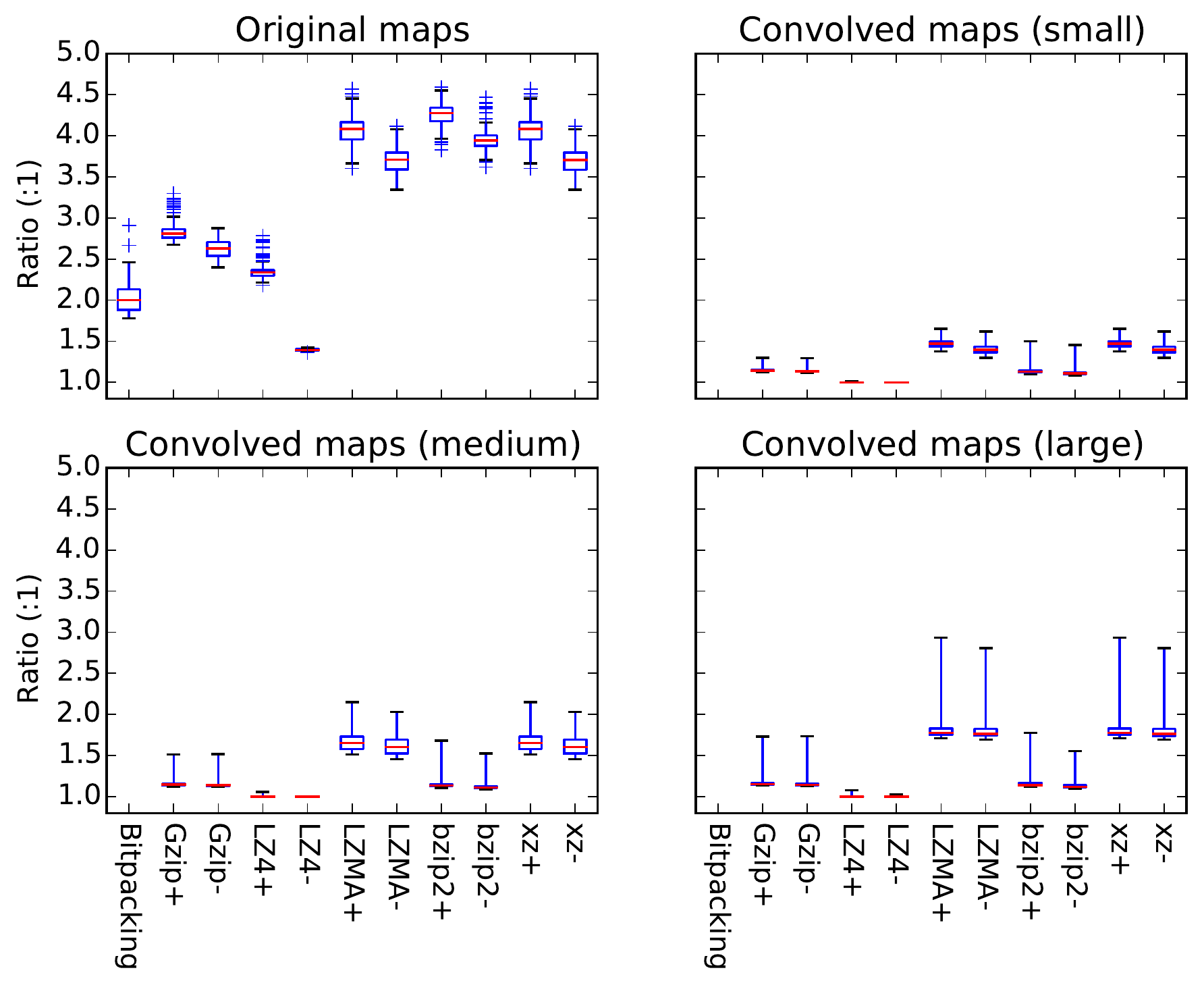}
\caption{Box and whiskers plots showing compression ratio statistics for 306 original maps (array of integers) and 306 maps convolved with $10^2$ (small), $100^2$ (medium) and $500^2$ (large)-pixel quasar profiles (array of floats), for different lossless compression software. The median (red line) is within the box bounded by the first and third quartiles range ($\mathtt{IQR}$ = $\mathtt{Q3}-\mathtt{Q1}$). The whiskers are $\mathtt{Q1-1.5 \times IQR}$ and $\mathtt{Q3+1.5 \times IQR}$. Beyond the whiskers, values are considered outliers and are plotted as crosses. $\texttt{Bitpacking}$ is only applicable on integer data. Therefore, there is no result relative to this technique for the convolved maps. For all other techniques, a plus (+) means that the technique was executed with the highest compression ratio flag (generally slower), while minus (-) used the lowest compression ratio flag (generally faster).}
\label{fig::lossless_ratio}
\end{figure*}

Table \ref{table::lossless_results} shows the mean results for median compression time (\emph{ctime}), decompression time (\emph{dtime}), and compression ratio (\emph{ratio}) for each map type (original and convolved maps) we used from GERLUMPH. The global compression ratio results are plotted in Figure \ref{fig::lossless_ratio}.

\begin{table*}[ht]
  \centering
  \caption{List of GERLUMPH parameter space's mean values: median compression time (ctime) in seconds, median decompression time (dtime) in seconds and compression ratio (ratio; \#:1) for original and convolved maps. A plus (+) means that the technique was executed with the highest compression ratio flag (generally slower), while minus (-) indicates the lowest compression ratio flag (generally faster). Bit packing is a technique only applicable for original maps (integer values). Bold indicates best results for a given map type.}
  \footnotesize\setlength{\tabcolsep}{6.3pt}
    \begin{tabular}{cc|cccccccccccc}     
        \toprule
&  & \multicolumn{2}{c}{\textbf{bzip2}} & \multicolumn{2}{c}{\textbf{Gzip}} & \multicolumn{2}{c}{\textbf{LZ4}} & \multicolumn{2}{c}{\textbf{LZMA}} & \multicolumn{2}{c}{\textbf{xz}} & \textbf{bitpacking} \\
&  & - & + & - & + & - & + & - & + & - & + &    \\
       \midrule     
\multirow{3}{*}{Original map}		& ctime & 44.91 & 44.27 & 12.50 & 507.44 & \textbf{2.53} & 33.16 & 53.78 & 245.78 & 54.47 & 247.50 & 6.39 \\
								& dtime & 15.05 & 19.46 & 6.34 & 5.12 & 1.67 & \textbf{1.24} & 16.06 & 16.53 & 16.91 & 17.44 & 6.12 \\
								& ratio & 3.94 & \textbf{4.28} & 2.63 & 2.81 & 1.39 & 2.34 & 3.71 & 4.08 & 3.71 & 4.08 & 2.00 \\
       \midrule     								
\multirow{3}{*}{$10^2$-pixel quasar profile} 		& ctime & 85.59 & 86.62 & 25.86 & 31.63 & \textbf{1.06} & 17.96 & 86.05 & 127.12 & 86.84 & 128.22 & ---\\
								& dtime & 37.18 & 38.79 & 6.67 & 6.27 & \textbf{0.99} & \textbf{0.99} & 37.88 & 38.26 & 38.55 & 39.04 & --- \\
								& ratio & 1.11 & 1.13 & 1.13 & 1.14 & 1.00 & 1.00 & 1.40 & \textbf{1.47} & 1.40 & \textbf{1.47} & --- \\
       \midrule     					
\multirow{3}{*}{$100^2$-pixel quasar profile} 		& ctime & 72.12 & 74.34 & 23.03 & 27.34 & \textbf{1.09} & 16.98 & 68.86 & 106.22 & 69.30 & 107.03 & --- \\
								& dtime & 33.35 & 35.59 & 5.94 & 5.61 & \textbf{1.05} & 1.06 & 26.79 & 26.82 & 27.35 & 27.46 & --- \\
								& ratio & 1.11 & 1.13 & 1.14 & 1.15 & 1.00 & 1.00 & 1.60 & \textbf{1.65} & 1.60 & \textbf{1.65} & ---  \\
       \midrule     										
\multirow{3}{*}{$500^2$-pixel quasar profile} 		& ctime & 65.55 & 67.11 & 20.43 & 24.36 & \textbf{0.94} & 14.56 & 60.44 & 97.07 & 60.88 & 97.68 & --- \\
								& dtime & 29.98 & 32.09 & 5.43 & 5.15 & \textbf{0.91} & 0.91 & 23.56 & 23.69 & 24.12 & 24.28 & --- \\
								& ratio & 1.12 & 1.14 & 1.14 & 1.15 & 1.00 & 1.00 & 1.76 & \textbf{1.78} & 1.76 & \textbf{1.78} & --- \\
        \bottomrule
    \end{tabular}
  \label{table::lossless_results}
\end{table*}

\newcommand{\figurescale}{0.505}

A first observation we can make is that all techniques perform better on arrays of integers than on arrays of floats. We also see that as we convolve with the larger quasar profile, the compression ratio for the convolved maps (array of floats) increases but is generally much lower than for the original maps (array of integers). Overall the compression and decompression times decrease as the size of the quasar profile is increased.


From the table and figures, we can draw several conclusions about the lossless compression software we evaluated. First of all, $\texttt{LZMA}$ and $\texttt{xz}$ provide the best compression ratio on average for convolved maps (floats), and $\texttt{bzip2}$ delivered the best compression ratio on average for original maps (integers). These three techniques are slow relative to the other techniques tested. $\texttt{xz}$ is slightly slower than $\texttt{LZMA}$. $\texttt{LZ4}$ and $\texttt{bitpacking}$ deliver lower compression ratio than their evaluated rivals, but are much faster than their tested counterparts. There is a tradeoff between speed and compression ratio; however, this gap is much smaller for $\texttt{bzip2}$. 

All of these compression techniques were faster at decompression than at compression, which is important for archiving purposes as one only needs to compress once, but may need to decompress many times. An important point to note is that $\texttt{LZ4}$ only compressed the integers files (where the mean compression ratio for convolved maps is $1$:$1$). Therefore, even if it was technically the fastest technique, it was practically useless for convolved maps.

Such lossless results would not enable the storage of all the expected data of the convolution phase of GERLUMPH. As discussed in section \ref{subsection::gerlumph}, the expected amount of generated data in the convolution phase is about 2.7 PB. Given the results for convolved maps, such lossless compression would only compression to the order of the petabyte, still exceeding the current available storage space for GERLUMPH of $\approx 50$ TB on gSTAR. 

%
%
\section{Lossy data compression experiment}
\label{section::lossyExperiment}
The lossy JPEG2000 standard has been evaluated in consideration of its potential as a future astronomy standard. As an extension of the work of \citet{kitaeff-2012}, \citet{Peters-2014}, and \citet[this issue]{Kitaeff-2014}, we evaluate the lossy JPEG2000 format for original and convolved magnification maps. 

\subsection{Evaluated software}
$\mathtt{KERLUMPH}$\footnote{Available at http://supercomputing.swin.edu.au/projects/kerlumph/}, the GERLUMPH extension of the Kakadu Software Development Kit (KDU, version 7.4)\footnote{http://kakadusoftware.com}, has been developed to take advantage of the image-like maps of GERLUMPH. It converts a binary file into a compressed $\mathtt{jp2}$ file \citep{Adams01thejpeg-2000}. An interesting feature of the $\mathtt{jp2}$ format is the availability of inner customizable metadata about the map. $\mathtt{KERLUMPH}$ enables lossy compression for both integer and real type of values, and to go back from $\mathtt{.jp2}$ to $\mathtt{.bin}$ in order to enable work with raw-like data. A low level description of the standards is beyond the scope of this paper, instead we refer the reader to \citet{taubman-2005} and \citet{li-2003} for explanations on the JPEG2000 standard and its related mathematics.

A $\texttt{KERLUMPH}$ user needs to supply the input file, the output file name ($\texttt{jp2}$ file), the dimensions of the input map, and the data type (32-bit integer or 32-bit float). It is also possible to supply other JPEG2000 parameters such as those tested in this paper (Qstep and Clevels) in order to overwrite the default parameter value. The application then evaluates the range of the input data, converts the data to floats (required for integers values), and scales it to the [-0.5, 0.5] range required for the lossy JPEG2000 compression. The original range and the data type are saved into the $\texttt{jp2}$ file header. Decompression inverts the compression process and returns a binary file with the original data type and range.

\subsection{Methodology}
\label{section::method_lossy}
The JPEG2000 parameter space can be overwhelming if every combination of compression parameters was to be tested. Table \ref{table::jpeg2000params} shows the list of parameters taken from the core processing system, Part 1 of the JPEG2000 standard \citep{JPEG2000-part1}. 

\begin{table}
  \centering
    \caption{Parameters in Part 1 of the JPEG2000 Standard, ordered as encountered in the encoder. The two parameters we investigated are highlighted.}
    \begin{tabular}{cl}
        \toprule
        & Parameter  \\
        \midrule
1. & Reconstructed image bit depth \\
2. & Tile size \\
3. & Color space \\
4. & Reversible or irreversible transform \\
\textbf{\emph{5.}} & \textbf{\emph{Number of wavelet transform levels}} \\ 
6. & Precinct size \\
7. & Code-block size \\
\textbf{\emph{8.}} & \textbf{\emph{Coefficient quantization step size}} \\
9. & Perceptual weights \\
10. & Block coding parameters: \\
&(a) Magnitude refinement coding method \\
&(b) MQ code termination method \\
11. & Progression order \\
12. & Number of quality layers \\
13. & Region of interest coding method \\
     \bottomrule
    \end{tabular}
  \label{table::jpeg2000params}
\end{table}

Given the $\kappa,\gamma$ parameter space of GERLUMPH that we are covering, testing all possible parameters and combinations of compression parameters would 
be extremely time consuming. Instead, as we are working with data that approximates grey scale still-image maps, we selected 2 parameters dictated by part 1 of the JPEG2000 standard (highlighted in Table \ref{table::jpeg2000params}). These parameters are:

\begin{enumerate}
	\item The coefficient quantization step size (\emph{Qstep}), which discretises the wavelet coefficient values. It enables a trade-off between compressed image quality and encoding efficiency \citep{Clark2008}.
	\item The number of levels (\emph{Clevels}) in the Discrete Wavelet Transform tree, which influences the wavelet domain before quantization and encoding. Increasing the number of DWT levels let examine the lower frequencies at increasingly finer resolution, packing more energy into fewer wavelet coefficients and leading to the expectation that compression performance improves as the number of levels increases \citep{Clark2008}.
	
\end{enumerate}

\begin{table}
  \centering
    \caption{JPEG2000 parameter space explored for each GERLUMPH map.}
    \begin{tabular}{ccccc}
        \toprule
		\textbf{Parameter} & \textbf{Default} & \textbf{Start} & \textbf{End} & \textbf{Step}  \\
        \midrule
Qstep  & $1/256$ & $10^{-7}$ & $\approx2$ & $\times \sqrt[4]{2}$\\
Clevels & 5 & 5 & 32 & +27\\
        \bottomrule
    \end{tabular}
  \label{table::jp2params}
\end{table}

\citet{Peters-2014} showed that the code block size and precincts size had no effect on both compression and soundness of the spectral cube data. Therefore, we have bypassed these parameters for this evaluation. Another difference in our approach is that we have not used $Clevels$ by itself with all other parameters set to default as in \citet{Peters-2014}. Instead, we have combined $Qstep$ with $Clevels$, with the hypothesis that compression performance would improve as the wavelet decomposition levels increase for a similar quantization step size. The explored parameter space is displayed in Table \ref{table::jp2params}. 

One can observe that our choices for the \emph{Qstep} parameter vary from those used in \citet{Peters-2014}, which started at $10^{-6}$ and ended at $0.01$. As our maps showed different order of details depending on the type (original or convolved), and depending on the quasar profile size used in the convolution phase, we used a broader range to let us find optimal values for our different map type. Our search for optimal compression is depicted in Algorithm \ref{algo::find_optima}. 

\begin{algorithm}
\DontPrintSemicolon
\KwData{$Map$, $Qsteps$ (sorted array based on Table \ref{table::jp2params})}
\KwResult{Optimal results for $Clevels \in \{5,32\}$}
\Begin{
	\ForEach{$Clevels \in \{5,32\}$}{
		Initialise $ratio_{best}, rmse_{best},qstep_{best}, previous_m$ \;
		$l \longleftarrow 0$,  $r \longleftarrow |Qsteps|-1$\;

	    	\While{$l \leq r$}{
			$m \longleftarrow \lfloor{\frac{l+r}{2}}\rfloor$\;
			$ratio, rmse \longleftarrow$ CompressAndCompare($map$, $Qsteps[m]$, $Clevels$)\;
			\If{$previous_m \neq m$}{
				\If{$rmse \leq 0.01$}{
					$ratio_{best}  \longleftarrow ratio$\;
					$rmse_{best}\longleftarrow rmse$\;
					$qstep_{best} \longleftarrow Qsteps[m]$\;
			                 \If{$m > l$}{
			                        $r \longleftarrow m-1$\;
			                 }
				}
				\Else{
					$l \longleftarrow m+1$\;
				}
				$previous_m \longleftarrow m$\;
			}
			\Else{$break$\;}
		}
		Save($ratio_{best}$, $rmse_{best}$, $qstep_{best}$, $Clevels$)\;
	}
}
\caption{Find optima for a given map}
\label{algo::find_optima}
\end{algorithm}

All lossy experiments were conducted using Linux (CentOS release 6.6) on SGI C2110G-RP5 nodes containing 2 eight-core SandyBridge processors at 2.2 GHz, where each processor is 64-bit 95W Intel Xeon E5-2660, with 4 GB of RAM allocated.

\subsubsection {Criteria for evaluating lossy compression}
Since we used JPEG2000 as a lossy compression technique, we have to evaluate its effect on future analysis. The microlensing community uses two main analysis techniques: the single-epoch imaging technique \citep{Bate2008,Floyd2009} and the light curve technique \citep{Kochanek2004}. 

\begin{figure}[!htb]
\centering
\includegraphics[width=8.9cm]{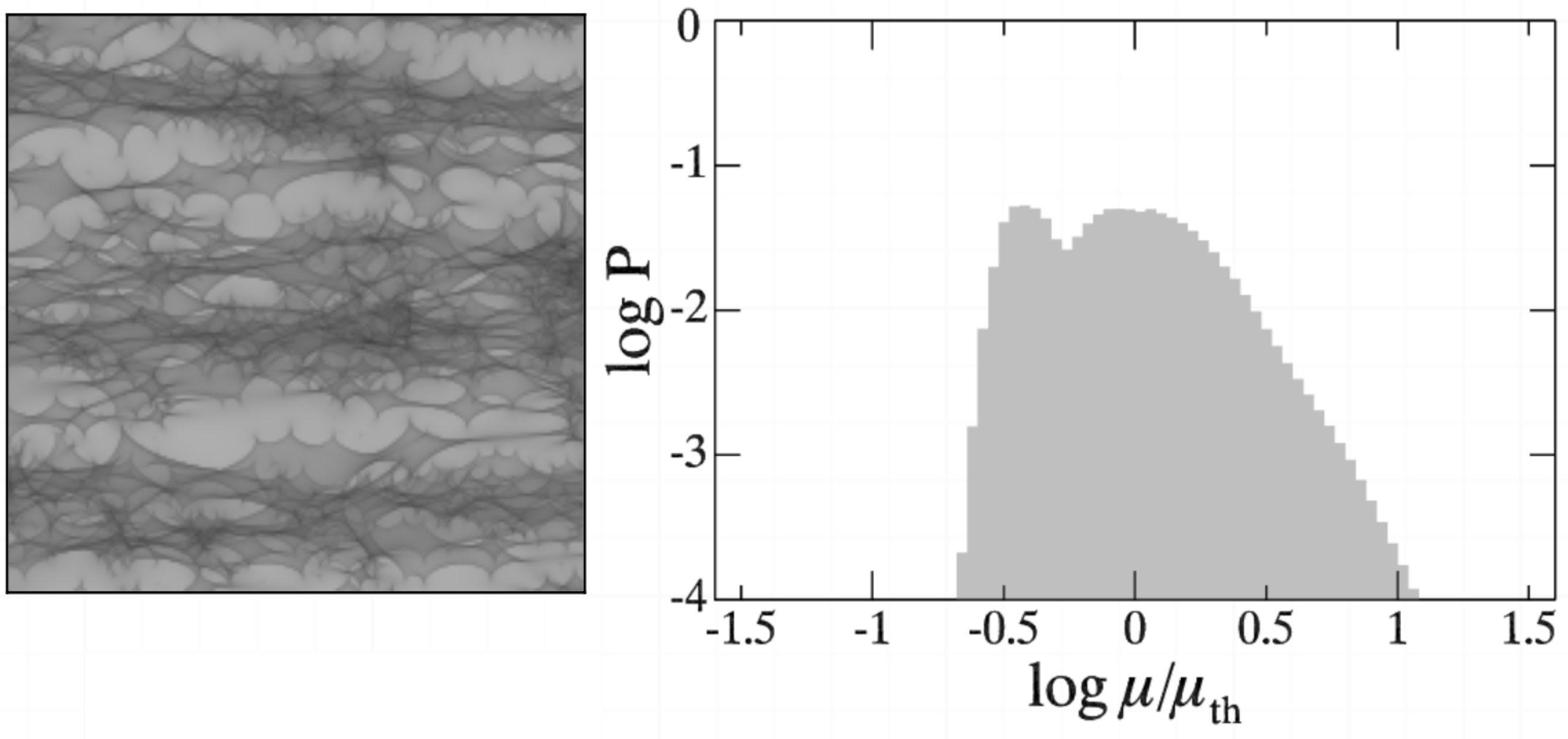}
\caption{The single-epoch imaging technique uses the MPD (right), an histogram of the count of each magnitude value in a magnification map (left) normalised by the total number of pixels. The map in this Figure is taken at $\kappa=0.52,\gamma=0.36$. Source: gerlumph.swin.edu.au.}
\label{fig::single-epoch}
\end{figure}

The single-epoch imaging technique uses the Magnification Probability Distribution (MPD) of the map. The MPD is a binned, normalized histogram of the count of each magnitude value, $mag$, in a map (Figure \ref{fig::single-epoch}):

\begin{equation}
mag = \log_{10} \frac{\mu_{i,j}}{\mu_{th}},
\end{equation}

with the local magnification value $\mu_{i,j}$ (Equation \ref{eq::local_magnification_value}) for pixel $i,j$, and the theoretical magnification $\mu_{th}$ for the macromodel: 

\begin{equation}
\mu_{th} = \frac{1}{(1-\kappa)^2-\gamma^2}.
\end{equation}


The light curve technique consists of extracting many light curves from the magnification map. A light curve is defined as a one dimensional array of neighbouring pixels. The value of each pixel of the light curve corresponds to the relative magnitude $\Delta mag$, where

\begin{equation}
\Delta mag = 2.5 \log_{10} \frac{\mu_{i,j}}{\mu_{th}}
\label{eq::relative_mag}
\end{equation}

for a given pixel at position $i,j$ in the magnification map. For each light curve extracted, the minimum, maximum, mean, and standard deviation of relative magnitude are calculated. In this work, we extracted light curves of $1 R_{Ein}$ in length and sampling length of 0.0025$R_{Ein}$. Using the values for $R_{Ein}$ and effective source velocity from \citet{KochanekSchechter2004} for Q2237+0305, such light curve would be $\approx15$ years long, sampled every $\approx13$ days. 


To evaluate the effect of compression, we have evaluated the root-mean-square-error (RMSE) to quantify the difference between the map before and after compression. The RMSE is defined as :

\begin{equation}
RMSE = \sqrt{\frac{1}{nm}\sum_{i=1}^{n}\sum_{j=1}^{m} (O_{i,j}-C_{i,j})^2},
\end{equation}

where $n$ is the length of the map in pixel, $m$ is the width of the map in pixel, $O_{i,j}$ is the relative magnitude value (Equation \ref{eq::relative_mag}) of pixel at position $i,j$ in the original map and $C_{i,j}$ is the relative magnitude value at pixel $i,j$ in the compressed distribution. 

To minimize the effect of compression on the analysis techniques described above, we have set a threshold on RMSE in order to obtain a near-lossless compression. As a guideline for RMSE, \citet{Poindexter2007} report a precision of 0.01, while \citet{Bate2008} report a precision in magnitude of 0.1. In this work, we consider an optimal near-lossless compression ratio such that $RMSE\leq0.01$.

\subsection{Results}
\label{section::results_lossy}

\begin{figure*}[!htb]
\centering
\includegraphics[width=18.4cm]{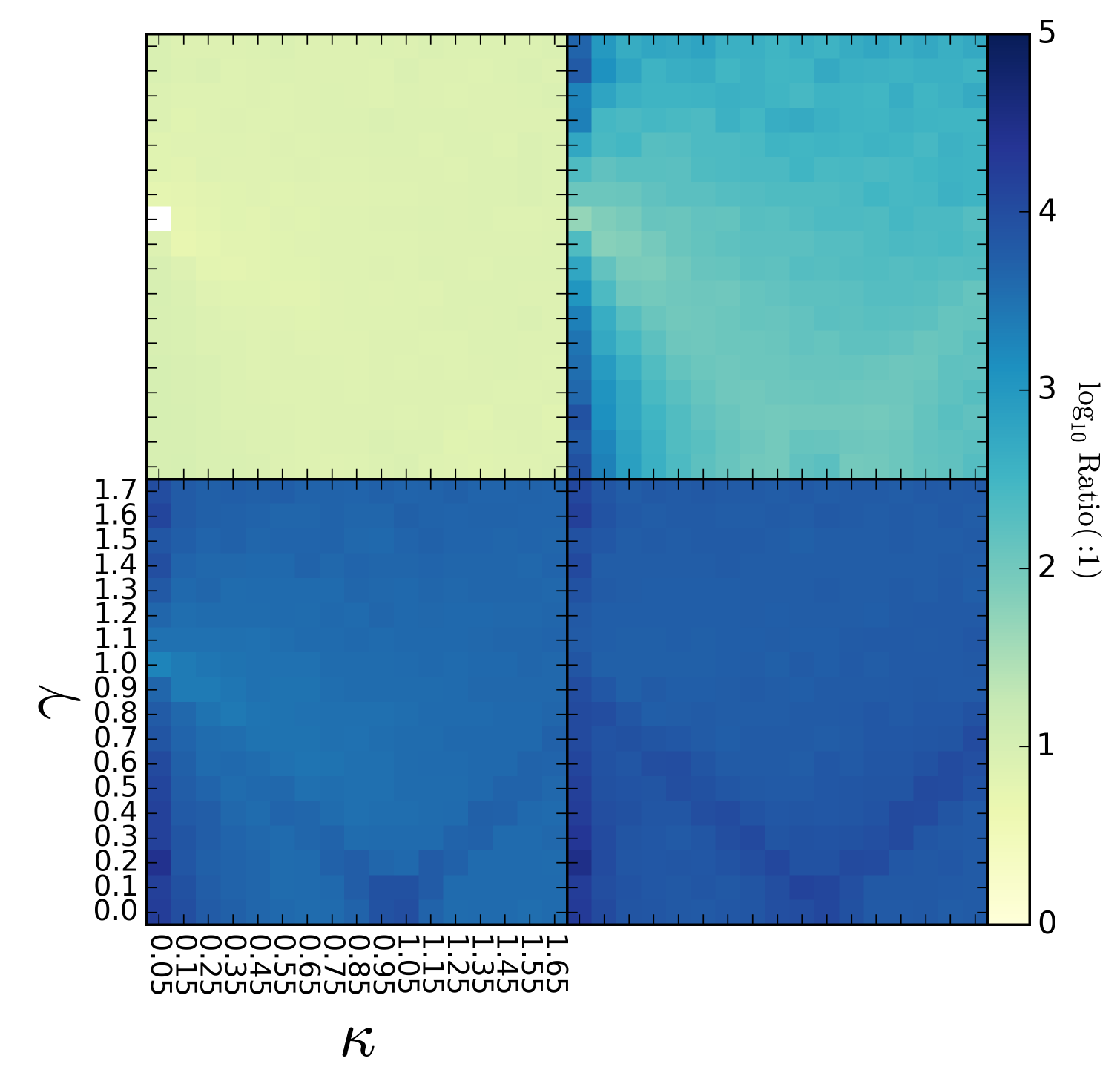}
\caption{Highest compression ratios obtained for $RMSE \leq 0.01$ using $Qstep$ and $Clevel=5$, plotted against the GERLUMPH parameter space (the shear $\gamma$ and the convergence $\kappa$). Each square represents a single magnification map. Quadrants show the original maps (upper left), and the maps convolved with a $10^2$-pixel (upper right), $100^2$-pixel (lower left), and $500^2$-pixel (lower right) quasar profile.}
\label{fig::lossy_ratio_5}
\end{figure*}

\begin{figure*}[!htb]
\centering
\includegraphics[width=18.4cm]{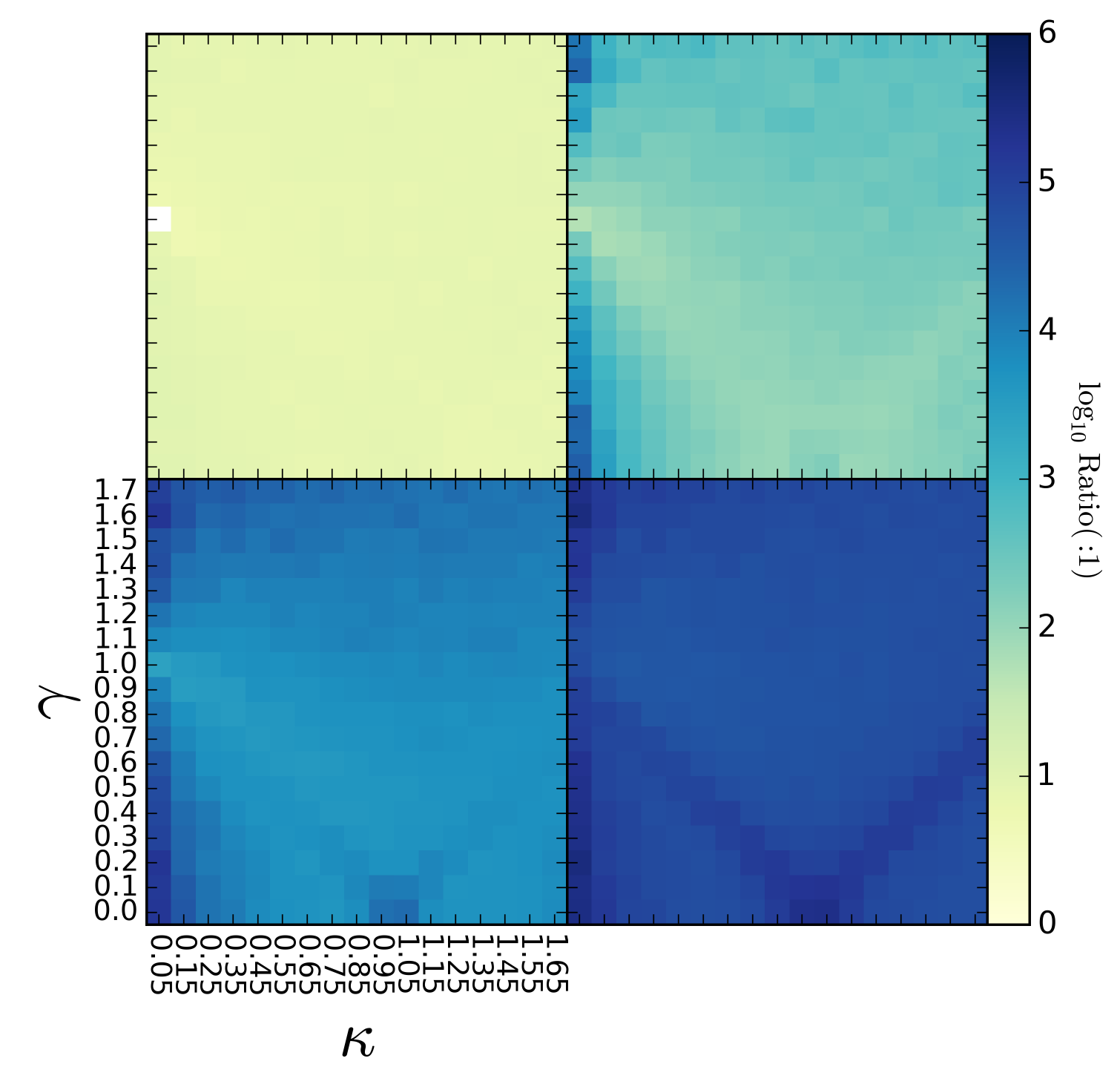}
\caption{Highest compression ratios obtained for $RMSE \leq 0.01$ using $Qstep$ and $Clevel=32$, plotted against the GERLUMPH parameter space (the shear $\gamma$ and the convergence $\kappa$). Each square represents a single magnification map. Quadrants show the original maps (upper left), and the maps convolved with a $10^2$-pixel (upper right), $100^2$-pixel (lower left), and $500^2$-pixel (lower right) quasar profile.}
\label{fig::lossy_ratio_32}
\end{figure*}

Table \ref{table::lossy_ratios} compares the minimum, maximum, mean and standard deviation in compression ratios obtained at $RMSE\leq0.01$ using $Clevel\in\{5,32\}$ for the different map types (original and convolved). We also show the individual compression ratios  against the GERLUMPH $\kappa,\gamma$ parameter space as obtained using $Clevel\in\{5,32\}$ in Figures \ref{fig::lossy_ratio_5} and \ref{fig::lossy_ratio_32} respectively. 

\begin{table*}[ht]
  \centering
  \caption{Comparison of compression ratio (\#:1) obtained using $Clevels \in \{5,32\}$, where $RMSE\leq0.01$, for all different GERLUMPH map types. Table shows the minimum, maximum, mean and standard deviation of all 306 maps per map type.}
  \footnotesize\setlength{\tabcolsep}{6.3pt}
    \begin{tabular}{c|cccccccc}     
        \toprule
 & \multicolumn{2}{c}{\textbf{Minimum}} & \multicolumn{2}{c}{\textbf{Maximum}} & \multicolumn{2}{c}{\textbf{Mean}}  & \multicolumn{2}{c}{\textbf{Standard deviation}} \\
$Clevels$ & 5 & 32 & 5 & 32 & 5 & 32 & 5 & 32  \\
       \midrule     
Original map & 5.14 & 5.14 & 10.00 & \textbf{10.01} & 7.70 & 7.70 & 0.72 & 0.72\\
$10^2$-pixel quasar profile & 48.55 & \textbf{48.88} & 8,429.24 & \textbf{30,121.51} & 420.30 & \textbf{732.24} & 947.65 & 3,017.25\\
$100^2$-pixel quasar profile & 1870.86 & \textbf{2,529.88} & 26,822.66 & \textbf{167,252.56} & 4,625.85 & \textbf{12,795.24} & 2,251.80 & 20,281.67\\
$500^2$-pixel quasar profile & 5,185.66 & \textbf{37,258.75} & 29,931.18 & \textbf{325,812.27} & 7,165.02 & \textbf{76,452.76} & 2,544.40 & 42,885.24\\
\bottomrule
    \end{tabular}
  \label{table::lossy_ratios}
\end{table*}

%

Only one of the original maps did not deliver a $RMSE \leq 0.01$ in the range of $Qstep$ evaluated (white square in the Figures at $\kappa=0.05,\gamma=1.0$). We did not broaden our search range as the lowest compression obtained within the range was smaller ($\approx$ 3:1) than the best lossless compression presented in Section \ref{section::losslessExperiment}. 

\begin{figure*}[!htb]
\setlength{\tabcolsep}{0.07cm}
\begin{tabular}{cc}
\multicolumn{2}{c}{\includegraphics[width=18.106cm]{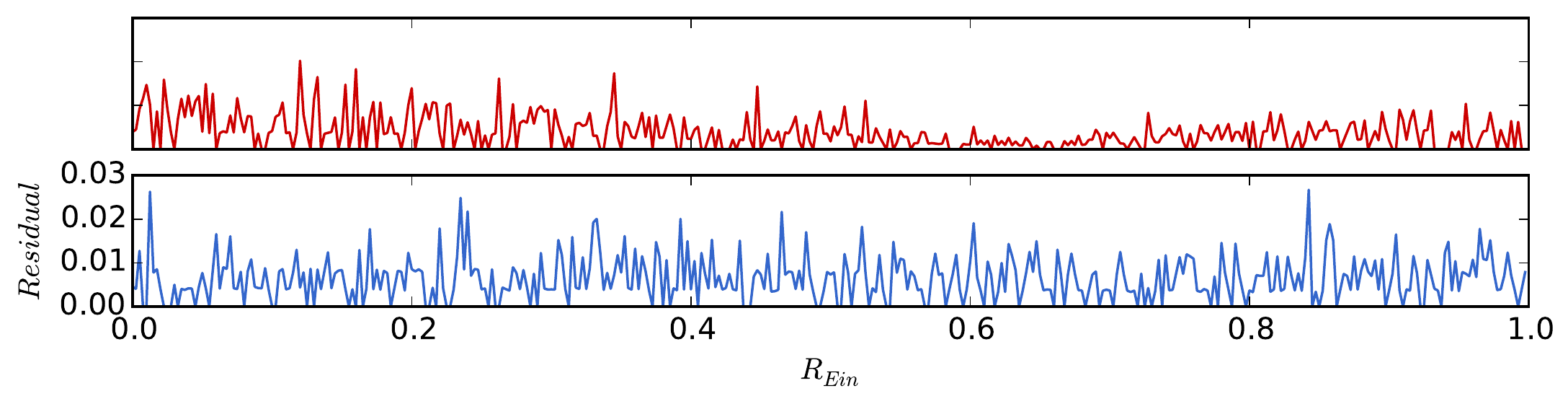}} \\
\includegraphics[width=9.053cm]{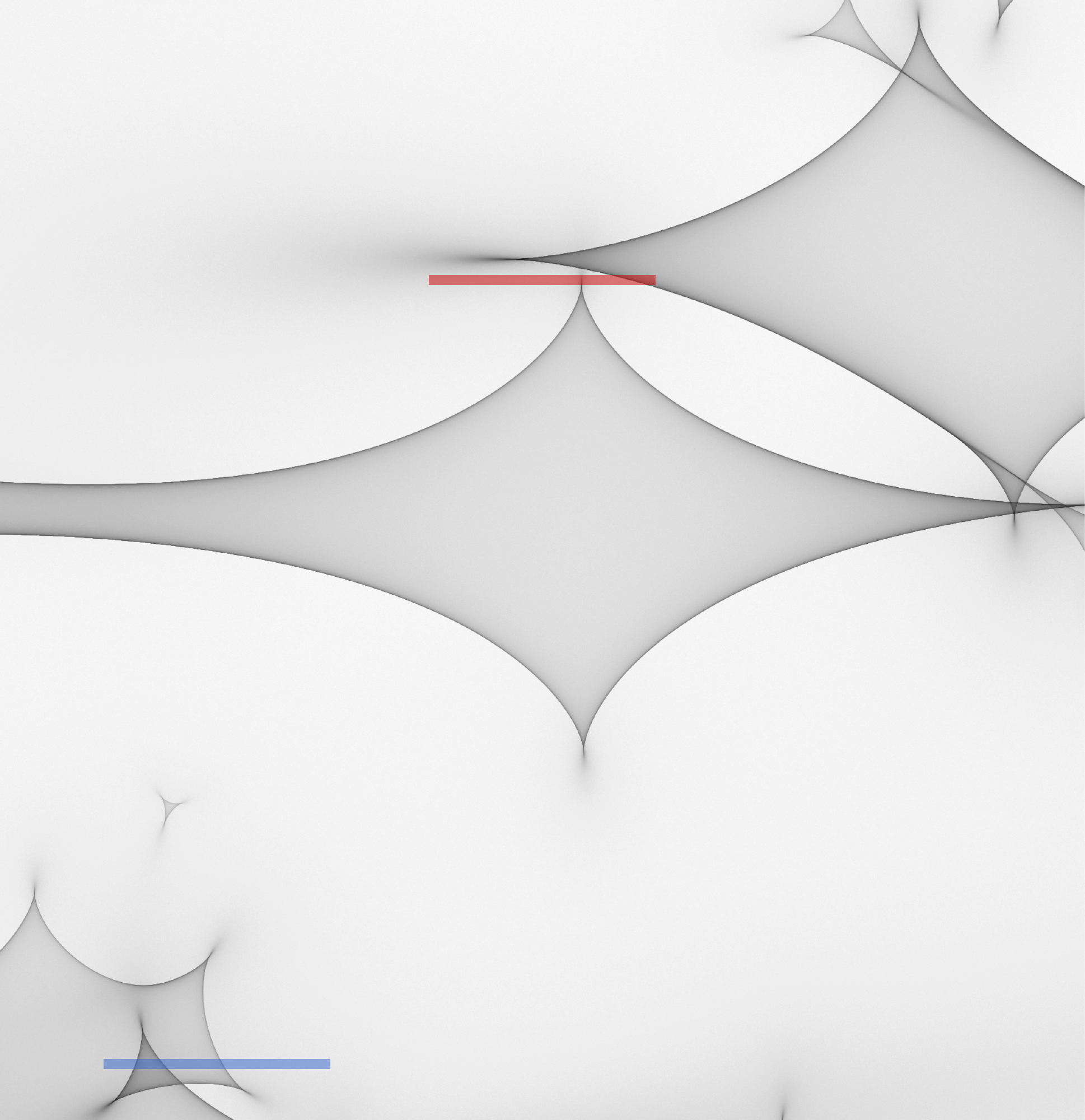} &
\includegraphics[width=9.053cm]{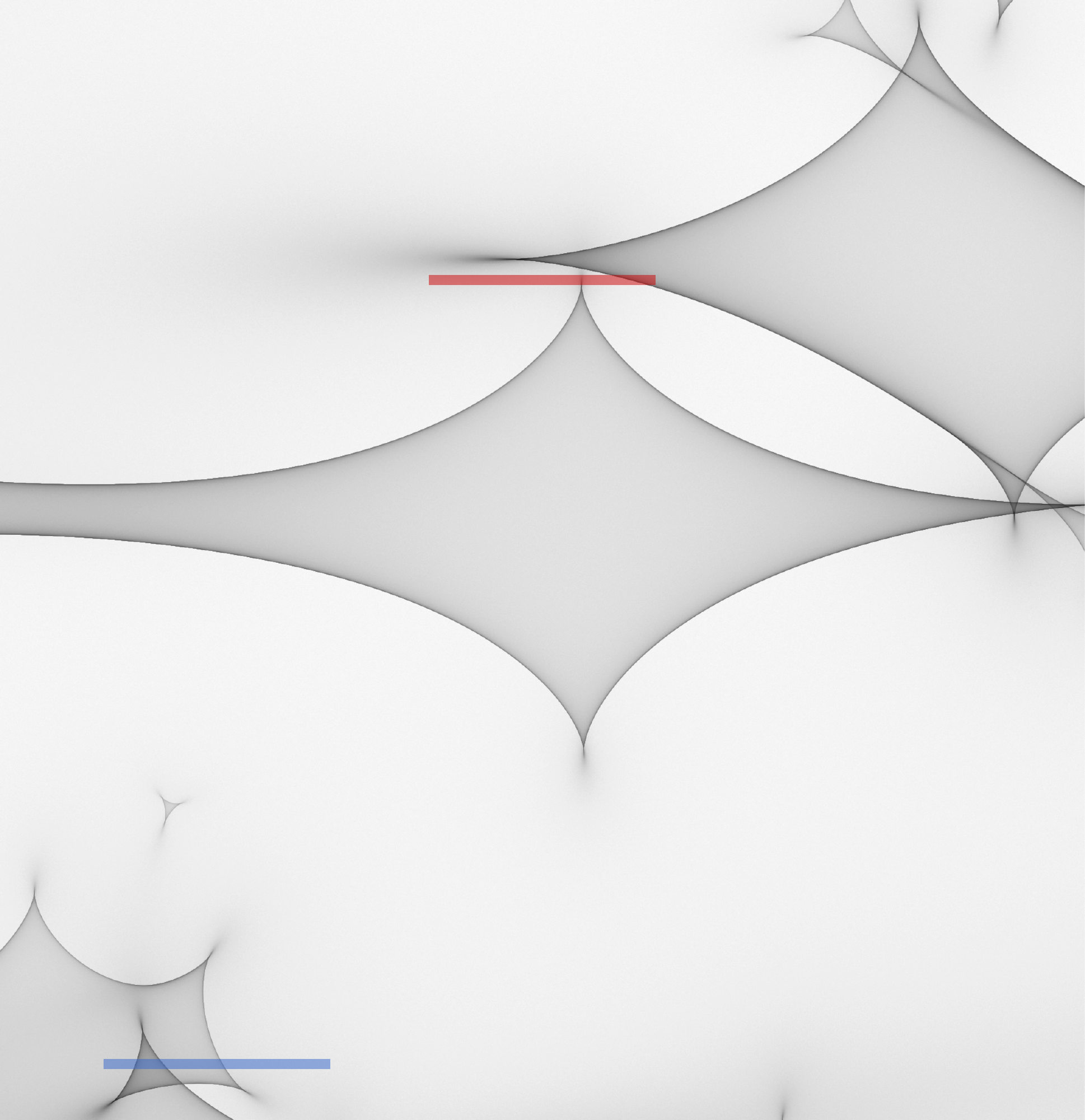}
\end{tabular}
\caption{Residual of two light curves of 1 $R_{ein}$ in length, extracted from an original map taken at $\kappa=0.25,\gamma=0.5$. Top panel shows the residual of two light curves (blue and red) extracted before and after compression. The same light curves are extracted from the map before (bottom left) and after (bottom right) compression (8:1) in order to evaluate the residual, at the locations depicted by the red and blue lines. The two maps in the bottom panel are $2000^2$-pixel close ups.}
\label{fig::lc_1}
\end{figure*}

\begin{figure*}[!htb]
\setlength{\tabcolsep}{0.07cm}
\begin{tabular}{cc}
\multicolumn{2}{c}{\includegraphics[width=18.106cm]{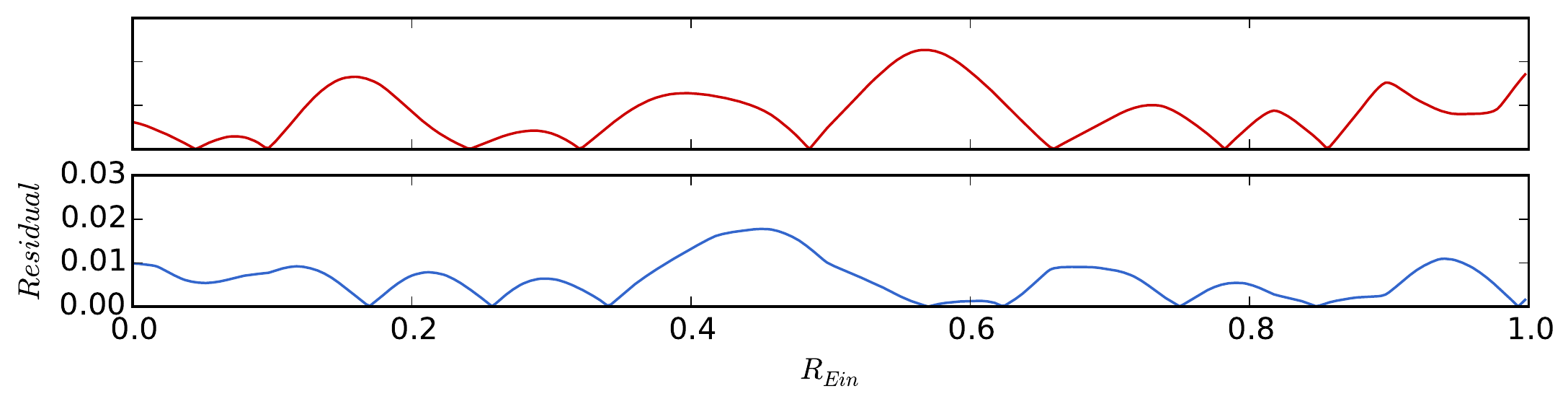}} \\
\includegraphics[width=9.053cm]{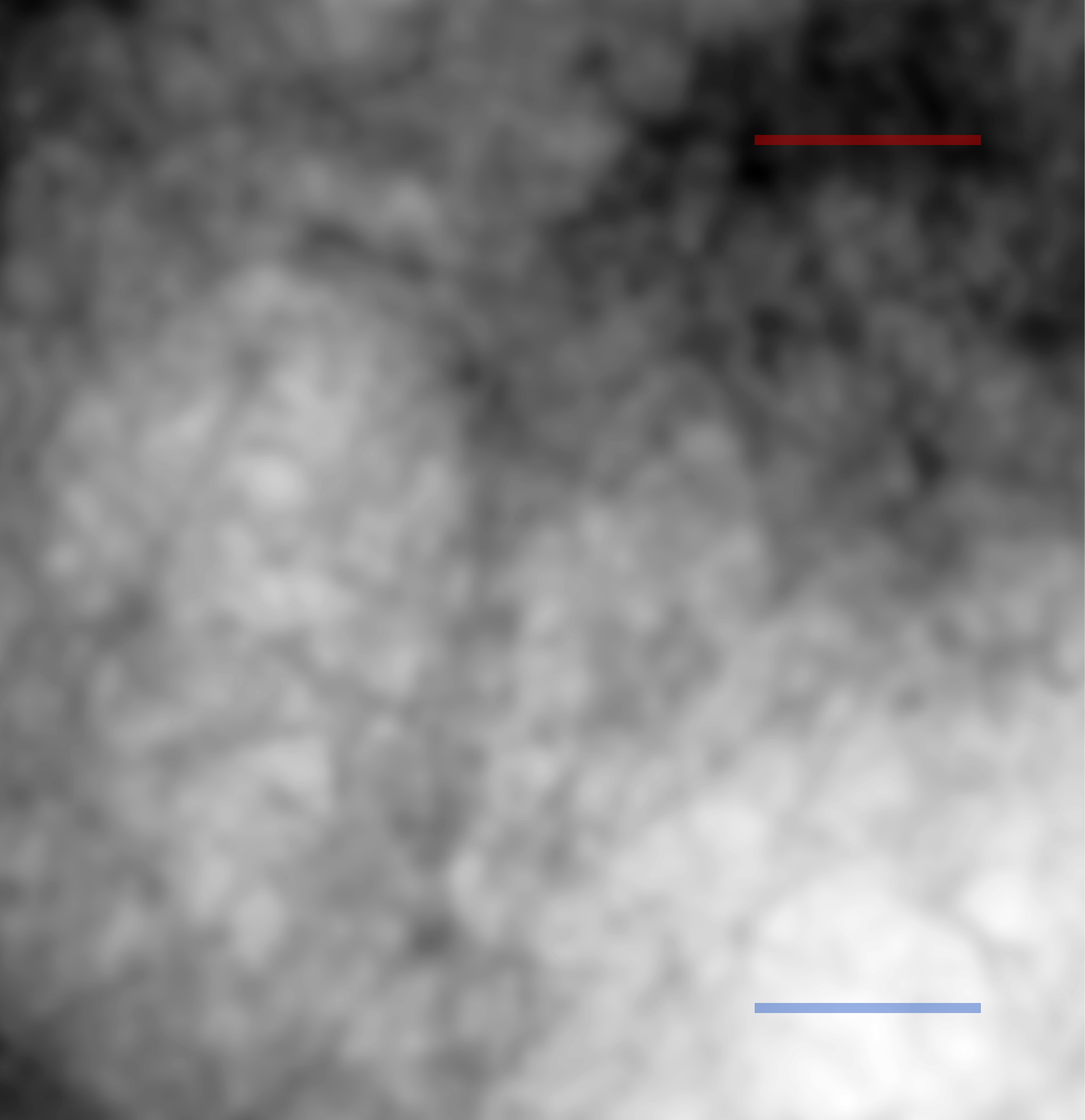} &
\includegraphics[width=9.053cm]{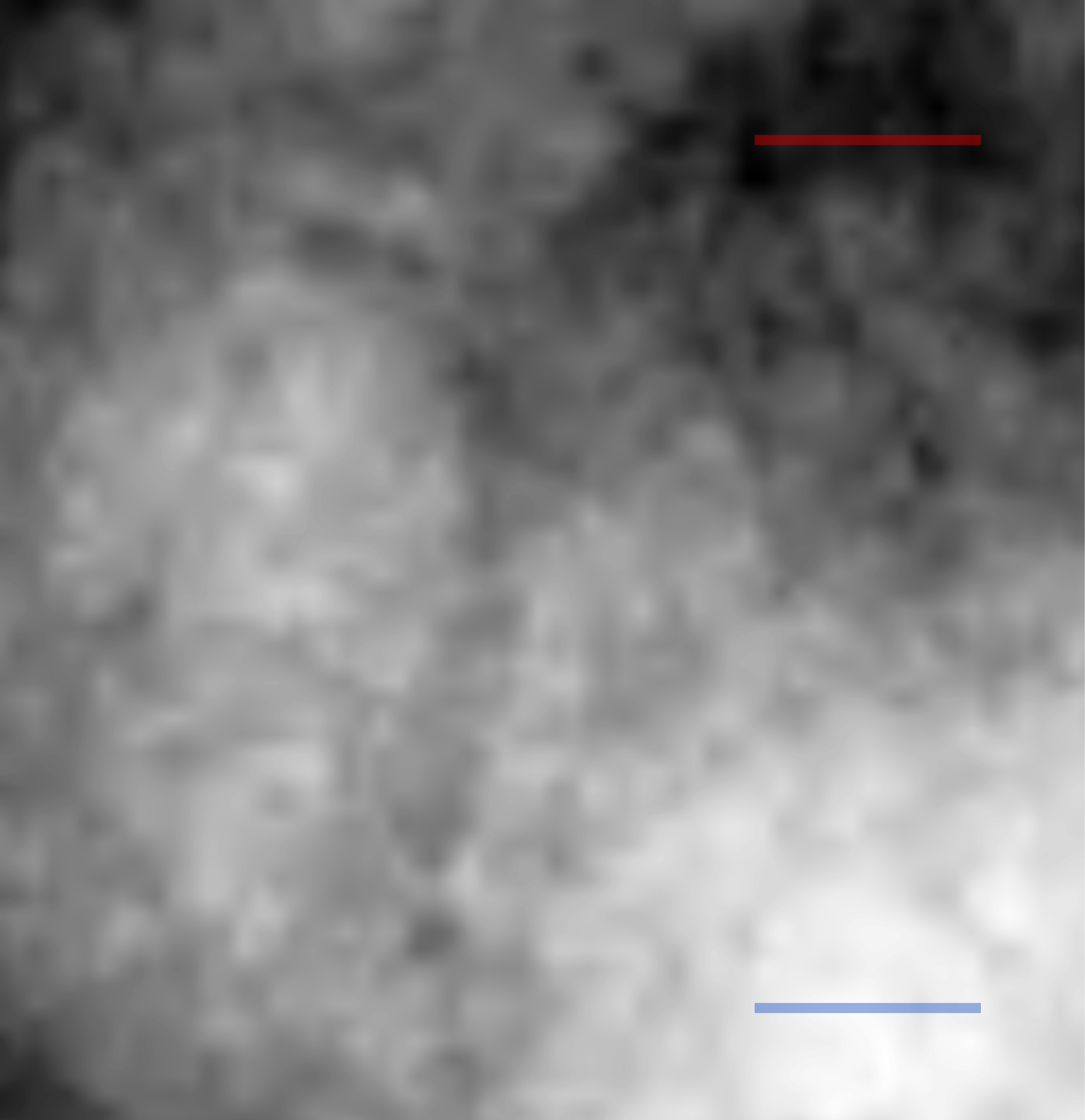}
\end{tabular}
\caption{Residual of two light curves of 1 $R_{ein}$ in length, extracted from a map convolved with a $100^2$-pixel quasar profile ($\kappa=0.95,\gamma=0.0$). Top panel shows the two residual light curves (blue and red). The same light curves are extracted from the map before (bottom left) and after (bottom right) compression (17,271:1) in order to evaluate the residual, at the locations depicted by the red and blue lines. The two maps in the bottom panel are $2000^2$-pixel close ups.}
\label{fig::lc_2}
\end{figure*}

In Figures \ref{fig::lc_1} and \ref{fig::lc_2}, we show the results of extracting two distinct light curves from an original map and a map convolved with a $100^2$-pixel quasar profile. Each light curve is depicted using a different colour. The figures show a close up of $2000^2$-pixel region for both the map before (lower left panel) and after compression (lower right panel). We extracted light curves at the same coordinate in both maps and evaluated the residual pixel by pixel. The residual of each light curves is shown in the top panel of each figure. As expected from using the RMSE as a threshold for the whole map, most subsampled pixels in the residuals are $\leq0.01$. 

\begin{figure*}[!htb]
\setlength{\tabcolsep}{0.07cm}
\begin{tabular}{cc}
\includegraphics[width=9.053cm]{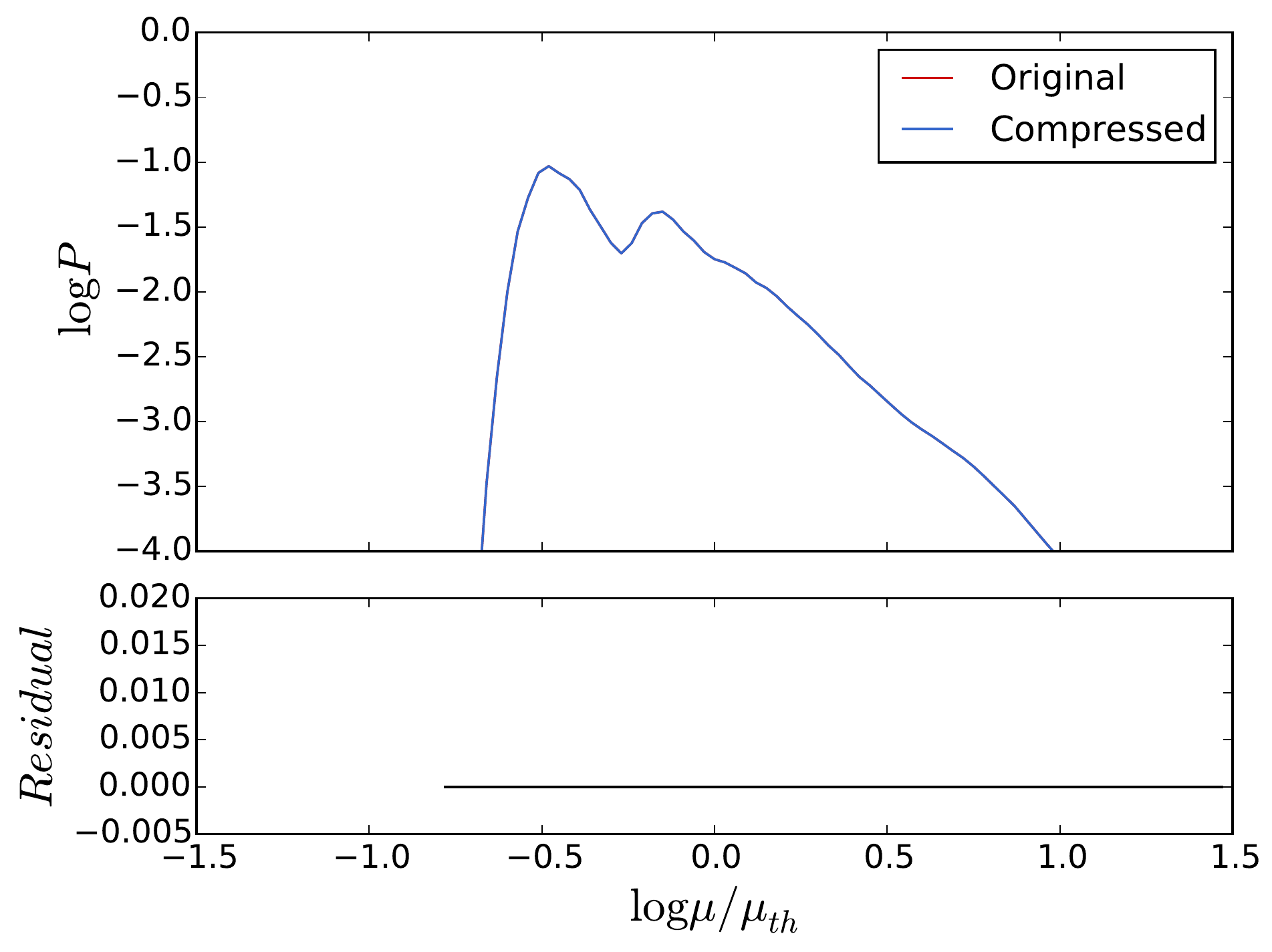} & \includegraphics[width=9.053cm]{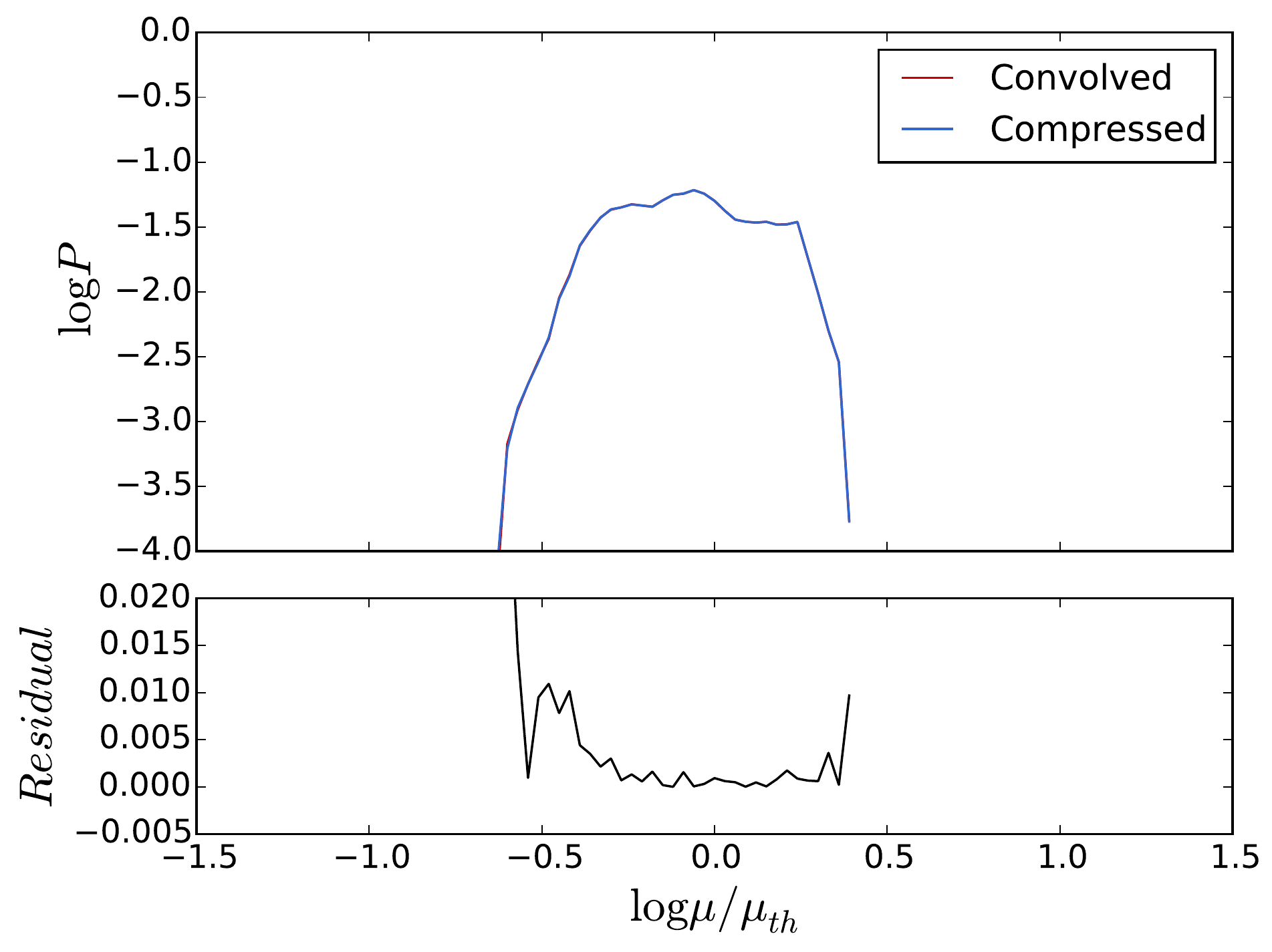}
\end{tabular}
\caption{Residual of two MPDs from an original map (left,  $\kappa=0.25,\gamma=0.5$) and a map convolved with a $100^2$-pixel quasar profile (right, $\kappa=0.95,\gamma=0.0$). These are the same maps used in Figures \ref{fig::lc_1} and \ref{fig::lc_2} respectively. Top panels show the two overlapping MPDs evaluated before and after compression (8:1 and 17,271:1 respectively) and bottom panels show the residual.}
\label{fig::MPD}
\end{figure*}

Similarly, we plot the MPD for the same maps in Figure \ref{fig::MPD}. The MPD of the original map stayed intact while showing limited differences for the map convolved with a $100^2$-pixel quasar profile. Most of the differences are found in demagnification, which would not have much consequences on analysis as an image that is demagnified would not likely be noticed anyway.

\begin{figure*}[!htb]
\centering
\includegraphics[width=18.4cm]{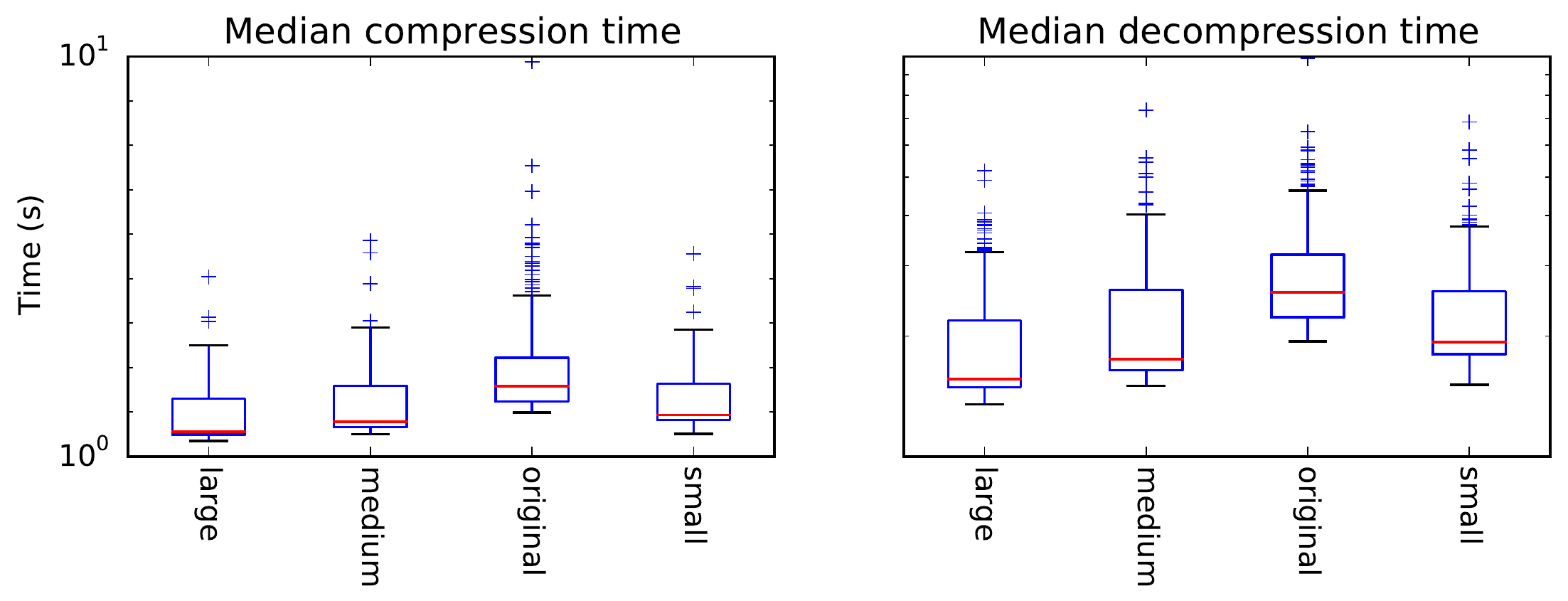}
\caption{Median compression time (left) and median decompression time (right) of three runs of compression obtained for optimal results shown in figure \ref{fig::lossy_ratio_32}. Maps convolved with $10^2$, $100^2$ and $500^2$ pixels quasar profile (array of floats) are marked with labels \emph{small}, \emph{medium} and \emph{large} respectively, and original maps (array of integers) are labeled \emph{original}.}
\label{fig::time_lossy}
\end{figure*}

An appealing feature of building $\mathtt{KERLUMPH}$ from the KDU is the speed delivered through its multi-threaded processing ability. The median compression time (s) and median decompression time (s) of three runs of compression are shown in Figure \ref{fig::time_lossy}. All compression and decompression was executed in under 10 seconds per map, while having median and mean time under 3 seconds (Table \ref{table::lossy_time}) on gSTAR.

\begin{table*}[ht]
  \centering
  \caption{List of GERLUMPH parameter space's mean and median time: median compression time (ctime) in seconds, median decompression time (dtime) in seconds and compression ratio (ratio; \#:1) for original maps and maps convolved with small, medium and large quasar profiles.}
  \footnotesize\setlength{\tabcolsep}{6.3pt}
    \begin{tabular}{c|cccccccc}     
        \toprule
& \multicolumn{2}{c}{\textbf{Original}} & \multicolumn{2}{c}{\textbf{Small}} & \multicolumn{2}{c}{\textbf{Medium}} & \multicolumn{2}{c}{\textbf{Large}} \\
& ctime & dtime & ctime & dtime & ctime & dtime & ctime & dtime  \\
       \midrule     
Distribution median & 2.58 & 2.57 & 1.94 & 1.93 & 1.78 & 1.75 & 1.56 & 1.56 \\
Distribution mean    & 2.88 & 2.85 & 2.21 & 2.23 & 2.14 & 2.18 & 1.85 & 1.89 \\        \bottomrule
    \end{tabular}
  \label{table::lossy_time}
\end{table*}



\section{Discussion}

As opposed to the lossless compression results which showed low compression ratios ($<$3:1 with a mean of 1.78:1 with $\texttt{LZMA}$ for the maps convolved with the largest quasar profile), the near-lossless compression results suggest that a high level of compression (up to 325,812:1, with mean of 76,452:1) can be applied to the convolved magnification maps with minimal impact on their future scientific use. Convolving maps with a quasar profile eliminates lots of fine details in the map --- the larger the quasar profile, the more fine details are eliminated (Figure \ref{fig::maps}). As mentioned in Section \ref{section::method_lossy}, increasing the \emph{Clevels} parameter lets JPEG2000 decompose the lower frequencies in the map at increasingly finer resolution. Convolved maps showed to be good candidates for such a technique. 

On the other hand, even if the near-lossless compression offered higher compression ratios than the lossless compression softwares we investigated (4.28:1 on average with $\texttt{bzip2}$), the compression ratio is still low (at most 10:1, with mean ratio of 7.70:1). If compression ratio is the most important issue, one should note that these results are higher than their lossless counterparts with limited error introduction.

As the original maps are the seeds for all the convolved maps, it is debatable that lossy compression should be applied to it, especially with the gain in compression ratio it provides. However, if it is indeed the path that is chosen, one can assume that minimal error will be introduced. Using the mean compression ratios, the lossless compression of original maps using $\texttt{bzip2}$ would shrink the storage requirement from $\approx27$ TB to $\approx6.3$ TB, while using near-lossless compression using JPEG2000 would shrink it to $\approx3.5$ TB. 

Given that we estimate the total amount of raw storage space required for the convolved maps to be two order of magnitude larger than the one of the original maps ($\approx2.7$ PB), the difference in compression ratio from lossless and near-lossless are very noticeable. If we do a rough estimate by using the mean compression ratio of all quasar profiles offered by $\texttt{LZMA}$ ($\approx1.6:1$), it would only shrink the storage requirement to $\approx1.7$ PB. Such compression ratio would not enable us to store all of the expected data from the next phase of GERLUMPH. However, if we do the same evaluation for the near-lossless compression mean ratio ($\approx$29,993.4:1), we would use only $\approx90$ GB for all convolved maps. Such storage space is already available within our resources.  Also, as the compressed files are much smaller in size than the original, its transmission requires less time and therefore saves bandwidth usage. 
 
Our case study results suggest that JPEG2000 could be suitable for other numerical datasets. In particular, well suited candidate are likely to be data structure such as gridded data or volumetric data. When approaching lossy data compression, one should keep in mind the intended purposes of the data to be compressed, and evaluate the effect of the loss on future analysis. In our case study, results suggest that such compression will still enable us to do our science.     

\section{Conclusion}
\label{section::conclusion}
We investigated a question relative to the petascale astronomy era: can data compression allow us to store large numerical simulation datasets? We used the GERLUMPH dataset comprising $\approx27$ TB of microlensing magnification maps as a case study. In the next phase of processing, each of the 70,000 GERLUMPH maps will be convolved with $\mathcal{O}(100)$ different quasar profiles. The dataset is expected to grow to $\approx2.7$ PB, way beyond the current storage space capacity of the project. 

We presented a comparative data compression study between several all-purpose (and one data-type specific) lossless compression software. We showed that such compression software delivered in best case a mean compression ratios of 4.28:1 for the original maps, and 1.47:1, 1.65:1 and 1.78:1 for maps convolved with $10^2$-pixel, $100^2$-pixel and $500^2$-pixel quasar profiles respectively. Such ratios do not provide a significant reduction in the required storage size. 

We also presented a follow-up study to that of \citet{Peters-2014} and \citet[this issue]{Kitaeff-2014} who investigated the JPEG2000 standard as a future astronomy data format standard. To do so, we evaluated its suitability for data generated in numerical simulation using near-lossless compression. We compared results of the two main techniques of quasar microlensing, the single-epoch imaging technique and the light curve technique, pre and post compression. We showed that JPEG2000 can deliver higher compression ratios (mean ratio of 7.7:1) for original maps than its lossless counterparts while keeping a low level of introduced errors. We also showed that for convolved maps, the compression ratios are much higher than in our lossless experiment (up to 325,812:1).

This study shows that JPEG2000 can be suitable for our case of simulated astronomical data. Such high compression ratios would enable us to comfortably compress 2.7 PB to less than a TB without corrupting the future science cases for which these data are meant for.

\section{Aknowledgements}
\label{section::aknowledgements}

This research was undertaken with the assistance of resources provided by gSTAR through the ASTAC scheme supported by the Australian Government's Education Investment Fund. Many thanks to Kakadu Software Ltd. and Dr. David Taubman for letting us use your KDU freely for this research.  This work was funded by the \emph{Fonds de recherche du Qu\'ebec --- Nature et technologies} (FRQNT) and Swinburne Research.




\bibliographystyle{model2-names}\biboptions{authoryear}
\bibliography{bibliography}

\begin{thebibliography}{64}
\expandafter\ifx\csname natexlab\endcsname\relax\def\natexlab#1{#1}\fi
\expandafter\ifx\csname url\endcsname\relax
  \def\url#1{\texttt{#1}}\fi
\expandafter\ifx\csname urlprefix\endcsname\relax\def\urlprefix{URL }\fi
\providecommand{\eprint}[2][]{\url{#2}}
\providecommand{\bibinfo}[2]{#2}
\ifx\xfnm\relax \def\xfnm[#1]{\unskip,\space#1}\fi
\bibitem[{Adams(2002)}]{Adams01thejpeg-2000}
\bibinfo{author}{Adams, M.D.}, \bibinfo{year}{2002}.
\newblock \bibinfo{title}{The jpeg-2000 still image compression standard}.
\bibitem[{{Ahn} et~al.(2012){Ahn}, {Alexandroff}, {Allende Prieto}, {Anderson},
  {Anderton}, {Andrews}, {Aubourg}, {Bailey}, {Balbinot}, {Barnes} \&
  et~al.}]{Ahn2012}
\bibinfo{author}{{Ahn}, C.P.}, \bibinfo{author}{{Alexandroff}, R.},
  \bibinfo{author}{{Allende Prieto}, C.}, \bibinfo{author}{{Anderson}, S.F.},
  \bibinfo{author}{{Anderton}, T.}, et~al., \bibinfo{year}{2012}.
\newblock \bibinfo{title}{{The Ninth Data Release of the Sloan Digital Sky
  Survey: First Spectroscopic Data from the SDSS-III Baryon Oscillation
  Spectroscopic Survey}}.
\newblock \bibinfo{journal}{The Astrophysical Journal Supplement}
  \bibinfo{volume}{203}, \bibinfo{pages}{21}.
\newblock \eprint{1207.7137}.
\bibitem[{{Anderson} et~al.(2011){Anderson}, {Alexov}, {B{\"a}hren},
  {Grie{\ss}meier}, {Wise} \& {Renting}}]{Anderson2011}
\bibinfo{author}{{Anderson}, K.}, \bibinfo{author}{{Alexov}, A.},
  \bibinfo{author}{{B{\"a}hren}, L.}, \bibinfo{author}{{Grie{\ss}meier}, J.M.},
  \bibinfo{author}{{Wise}, M.}, et~al., \bibinfo{year}{2011}.
\newblock \bibinfo{title}{{LOFAR and HDF5: Toward a New Radio Data Standard}},
  in: \bibinfo{editor}{{Evans}, I.N.}, \bibinfo{editor}{{Accomazzi}, A.},
  \bibinfo{editor}{{Mink}, D.J.}, \bibinfo{editor}{{Rots}, A.H.} (Eds.),
  \bibinfo{booktitle}{Astronomical Data Analysis Software and Systems XX},
  p.~\bibinfo{pages}{53}.
\newblock \eprint{1012.2266}.
\bibitem[{Andrianov(1984)}]{Andrianov1984}
\bibinfo{author}{Andrianov, S.A.}, \bibinfo{year}{1984}.
\newblock \bibinfo{title}{{Algorithms for the compression and coding of
  radio-astronomical data}}.
\newblock pp. \bibinfo{pages}{157--164}.
\bibitem[{{Bate} et~al.(2008){Bate}, {Floyd}, {Webster} \& {Wyithe}}]{Bate2008}
\bibinfo{author}{{Bate}, N.F.}, \bibinfo{author}{{Floyd}, D.J.E.},
  \bibinfo{author}{{Webster}, R.L.}, \bibinfo{author}{{Wyithe}, J.S.B.},
  \bibinfo{year}{2008}.
\newblock \bibinfo{title}{{A microlensing study of the accretion disc in the
  quasar MG 0414+0534}}.
\newblock \bibinfo{journal}{Monthly Notices of the Royal Astronomical Society}
  \bibinfo{volume}{391}, \bibinfo{pages}{1955--1960}.
\newblock \eprint{0810.1092}.
\bibitem[{Bentley et~al.(1986)Bentley, Sleator, Tarjan \& Wei}]{Bentley1986}
\bibinfo{author}{Bentley, J.L.}, \bibinfo{author}{Sleator, D.D.},
  \bibinfo{author}{Tarjan, R.E.}, \bibinfo{author}{Wei, V.K.},
  \bibinfo{year}{1986}.
\newblock \bibinfo{title}{A locally adaptive data compression scheme}.
\newblock \bibinfo{journal}{Communications of the ACM} \bibinfo{volume}{29},
  \bibinfo{pages}{320--330}.
\bibitem[{{Booth} et~al.(2009){Booth}, {de Blok}, {Jonas} \&
  {Fanaroff}}]{Booth2009}
\bibinfo{author}{{Booth}, R.S.}, \bibinfo{author}{{de Blok}, W.J.G.},
  \bibinfo{author}{{Jonas}, J.L.}, \bibinfo{author}{{Fanaroff}, B.},
  \bibinfo{year}{2009}.
\newblock \bibinfo{title}{{MeerKAT Key Project Science, Specifications, and
  Proposals}}.
\newblock \bibinfo{journal}{ArXiv e-prints} \eprint{0910.2935}.
\bibitem[{Broekema et~al.(2012)Broekema, Boonstra, Cabezas, Engbersen, Holties,
  Jelitto, Luijten, Maat, van Nieuwpoort, Nijboer, Romein \&
  Offrein}]{Broekema2012}
\bibinfo{author}{Broekema, P.C.}, \bibinfo{author}{Boonstra, A.J.},
  \bibinfo{author}{Cabezas, V.C.}, \bibinfo{author}{Engbersen, T.},
  \bibinfo{author}{Holties, H.}, et~al., \bibinfo{year}{2012}.
\newblock \bibinfo{title}{Dome: towards the astron \& ibm center for exascale
  technology}, in: \bibinfo{booktitle}{Proceedings of the 2012 workshop on
  High-Performance Computing for Astronomy}, \bibinfo{publisher}{ACM},
  \bibinfo{address}{New York, NY, USA}. pp. \bibinfo{pages}{1--4}.
\bibitem[{Burrows \& Wheeler(1994)}]{burrows-1994}
\bibinfo{author}{Burrows, M.}, \bibinfo{author}{Wheeler, D.J.},
  \bibinfo{year}{1994}.
\newblock \bibinfo{title}{A block-sorting lossless data compression algorithm}.
\newblock \bibinfo{type}{Technical Report}. Systems Research Center.
\bibitem[{{Cafforio} et~al.(1985){Cafforio}, {de Lotto}, {Rocca} \&
  {Savini}}]{Cafforio1985}
\bibinfo{author}{{Cafforio}, C.}, \bibinfo{author}{{de Lotto}, I.},
  \bibinfo{author}{{Rocca}, F.}, \bibinfo{author}{{Savini}, M.},
  \bibinfo{year}{1985}.
\newblock \bibinfo{title}{{Data Compression Techniques in Image Processing for
  Astronomy}}, in: \bibinfo{editor}{{di Gesu}, V.}, \bibinfo{editor}{{Scarsi},
  L.}, \bibinfo{editor}{{Crane}, P.}, \bibinfo{editor}{{Friedman}, J.H.},
  \bibinfo{editor}{{Levialdi}, S.} (Eds.), \bibinfo{booktitle}{Data Analysis in
  Astronomy}, p. \bibinfo{pages}{425}.
\bibitem[{{Clark}(2008)}]{Clark2008}
\bibinfo{author}{{Clark}, A.}, \bibinfo{year}{2008}.
\newblock \bibinfo{title}{{Streamlining digital signal processing: A tricks of
  the trade guidebook (R. Lyons, Ed.) [book review]}}.
\newblock \bibinfo{journal}{IEEE Signal Processing Magazine}
  \bibinfo{volume}{25}, \bibinfo{pages}{146--147}.
\bibitem[{{Fixsen} et~al.(2000){Fixsen}, {Hanisch}, {Mather},
  {Nieto-Santisteban}, {Offenberg}, {Sengupta} \& {Stockman}}]{Fixsen2000}
\bibinfo{author}{{Fixsen}, D.J.}, \bibinfo{author}{{Hanisch}, R.J.},
  \bibinfo{author}{{Mather}, J.C.}, \bibinfo{author}{{Nieto-Santisteban},
  M.A.}, \bibinfo{author}{{Offenberg}, J.D.}, et~al., \bibinfo{year}{2000}.
\newblock \bibinfo{title}{{Cosmic Ray Rejection and Data Compression for
  NGST}}, in: \bibinfo{editor}{{Manset}, N.}, \bibinfo{editor}{{Veillet}, C.},
  \bibinfo{editor}{{Crabtree}, D.} (Eds.), \bibinfo{booktitle}{Astronomical
  Data Analysis Software and Systems IX}, p. \bibinfo{pages}{539}.
\bibitem[{{Floyd} et~al.(2009){Floyd}, {Bate} \& {Webster}}]{Floyd2009}
\bibinfo{author}{{Floyd}, D.J.E.}, \bibinfo{author}{{Bate}, N.F.},
  \bibinfo{author}{{Webster}, R.L.}, \bibinfo{year}{2009}.
\newblock \bibinfo{title}{{The accretion disc in the quasar SDSS J0924+0219}}.
\newblock \bibinfo{journal}{Monthly Notices of the Royal Astronomical Society}
  \bibinfo{volume}{398}, \bibinfo{pages}{233--239}.
\newblock \eprint{0905.2651}.
\bibitem[{Frank(2002)}]{frank}
\bibinfo{author}{Frank, M.P.}, \bibinfo{year}{2002}.
\newblock \bibinfo{title}{The physical limits of computing}.
\newblock \bibinfo{journal}{Computing in Science and Engg.}
  \bibinfo{volume}{4}, \bibinfo{pages}{16--26}.
\bibitem[{{Gaudet} et~al.(2000){Gaudet}, {V{\'e}ran}, {Delisle} \&
  {Pirenne}}]{Gaudet2000}
\bibinfo{author}{{Gaudet}, S.}, \bibinfo{author}{{V{\'e}ran}, J.P.},
  \bibinfo{author}{{Delisle}, D.}, \bibinfo{author}{{Pirenne}, B.},
  \bibinfo{year}{2000}.
\newblock \bibinfo{title}{{Compression of Mosaic CCD Images with CompFITS2}},
  in: \bibinfo{editor}{{Manset}, N.}, \bibinfo{editor}{{Veillet}, C.},
  \bibinfo{editor}{{Crabtree}, D.} (Eds.), \bibinfo{booktitle}{Astronomical
  Data Analysis Software and Systems IX}, p. \bibinfo{pages}{547}.
\bibitem[{{Guzman} \& {Humphreys}(2010)}]{Guzman2010}
\bibinfo{author}{{Guzman}, J.C.}, \bibinfo{author}{{Humphreys}, B.},
  \bibinfo{year}{2010}.
\newblock \bibinfo{title}{{The Australian SKA Pathfinder (ASKAP) software
  architecture}}, in: \bibinfo{booktitle}{Society of Photo-Optical
  Instrumentation Engineers (SPIE) Conference Series}, p.~\bibinfo{pages}{1}.
\bibitem[{{Hassan} \& {Fluke}(2011)}]{HassanFluke2011}
\bibinfo{author}{{Hassan}, A.}, \bibinfo{author}{{Fluke}, C.J.},
  \bibinfo{year}{2011}.
\newblock \bibinfo{title}{{Scientific Visualization in Astronomy: Towards the
  Petascale Astronomy Era}}.
\newblock \bibinfo{journal}{Publications of the Astronomical Society of
  Australia} \bibinfo{volume}{28}, \bibinfo{pages}{150--170}.
\newblock \eprint{1102.5123}.
\bibitem[{{Hassan} et~al.(2012){Hassan}, {Fluke} \& {Barnes}}]{Hassan-2012}
\bibinfo{author}{{Hassan}, A.H.}, \bibinfo{author}{{Fluke}, C.J.},
  \bibinfo{author}{{Barnes}, D.G.}, \bibinfo{year}{2012}.
\newblock \bibinfo{title}{{A Distributed GPU-Based Framework for Real-Time 3D
  Volume Rendering of Large Astronomical Data Cubes}}.
\newblock \bibinfo{journal}{Publications of the Astronomical Society of
  Australia} \bibinfo{volume}{29}, \bibinfo{pages}{340--351}.
\newblock \eprint{1205.0282}.
\bibitem[{{Hassan} et~al.(2013){Hassan}, {Fluke}, {Barnes} \&
  {Kilborn}}]{Hassan-2013}
\bibinfo{author}{{Hassan}, A.H.}, \bibinfo{author}{{Fluke}, C.J.},
  \bibinfo{author}{{Barnes}, D.G.}, \bibinfo{author}{{Kilborn}, V.A.},
  \bibinfo{year}{2013}.
\newblock \bibinfo{title}{{Tera-scale astronomical data analysis and
  visualization}}.
\newblock \bibinfo{journal}{Monthly Notices of the Royal Astronomical Society}
  \bibinfo{volume}{429}, \bibinfo{pages}{2442--2455}.
\newblock \eprint{1211.4896}.
\bibitem[{Huffman(1952)}]{huffman}
\bibinfo{author}{Huffman, D.}, \bibinfo{year}{1952}.
\newblock \bibinfo{title}{{A Method for the Construction of Minimum-Redundancy
  Codes}}, in: \bibinfo{booktitle}{Proceedings of the I.R.E.}, pp.
  \bibinfo{pages}{1098--1102}.
\bibitem[{{IBM}(2013)}]{ibm-2013}
\bibinfo{author}{{IBM}}, \bibinfo{year}{2013}.
\newblock \bibinfo{title}{{S}quare {K}ilometer {A}rray: {U}ltimate {B}ig {D}ata
  {C}hallenge}.
\bibitem[{{ISO/IEC 15444-1:2000}(2000)}]{JPEG2000-part1}
\bibinfo{author}{{ISO/IEC 15444-1:2000}}, \bibinfo{year}{2000}.
\newblock \bibinfo{title}{{Information technology -- JPEG 2000 image coding
  system -- Part 1: Core coding system}}.
\newblock \bibinfo{type}{Technical Report}. {ISO/IEC}.
\bibitem[{{Kayser} et~al.(1986){Kayser}, {Refsdal} \& {Stabell}}]{Kayser1986}
\bibinfo{author}{{Kayser}, R.}, \bibinfo{author}{{Refsdal}, S.},
  \bibinfo{author}{{Stabell}, R.}, \bibinfo{year}{1986}.
\newblock \bibinfo{title}{{Astrophysical applications of gravitational
  micro-lensing}}.
\newblock \bibinfo{journal}{Astronomy and Astrophysics} \bibinfo{volume}{166},
  \bibinfo{pages}{36--52}.
\bibitem[{Kitaeff et~al.(2014)Kitaeff, Cannon, Wicenec \&
  Taubman}]{Kitaeff-2014}
\bibinfo{author}{Kitaeff, V.}, \bibinfo{author}{Cannon, A.},
  \bibinfo{author}{Wicenec, A.}, \bibinfo{author}{Taubman, D.},
  \bibinfo{year}{2014}.
\newblock \bibinfo{title}{{Astronomical imagery: Considerations for a
  contemporary approach with JPEG2000}}.
\newblock \bibinfo{journal}{Astronomy and Computing} \bibinfo{note}{(This
  issue)}.
\bibitem[{{Kitaeff} et~al.(2012){Kitaeff}, {Wu}, {Wicenec}, {Cannon} \&
  {Vinsen}}]{kitaeff-2012}
\bibinfo{author}{{Kitaeff}, V.V.}, \bibinfo{author}{{Wu}, C.},
  \bibinfo{author}{{Wicenec}, A.}, \bibinfo{author}{{Cannon}, A.D.},
  \bibinfo{author}{{Vinsen}, K.}, \bibinfo{year}{2012}.
\newblock \bibinfo{title}{{SkuareView: Client-Server Framework for Accessing
  Extremely Large Radio Astronomy Image Data}}.
\newblock \bibinfo{journal}{ArXiv e-prints} \eprint{1209.1877}.
\bibitem[{{Kochanek}(2004)}]{Kochanek2004}
\bibinfo{author}{{Kochanek}, C.S.}, \bibinfo{year}{2004}.
\newblock \bibinfo{title}{{Quantitative Interpretation of Quasar Microlensing
  Light Curves}}.
\newblock \bibinfo{journal}{The Astrophysical Journal} \bibinfo{volume}{605},
  \bibinfo{pages}{58--77}.
\newblock \eprint{astro-ph/0307422}.
\bibitem[{{Kochanek} \& {Schechter}(2004)}]{KochanekSchechter2004}
\bibinfo{author}{{Kochanek}, C.S.}, \bibinfo{author}{{Schechter}, P.L.},
  \bibinfo{year}{2004}.
\newblock \bibinfo{title}{{The Hubble Constant from Gravitational Lens Time
  Delays}}.
\newblock \bibinfo{journal}{Measuring and Modeling the Universe} ,
  \bibinfo{pages}{117}\eprint{astro-ph/0306040}.
\bibitem[{{Labrum} et~al.(1975){Labrum}, {McLean} \& {Wild}}]{Labrum1975}
\bibinfo{author}{{Labrum}, N.R.}, \bibinfo{author}{{McLean}, D.J.},
  \bibinfo{author}{{Wild}, J.P.}, \bibinfo{year}{1975}.
\newblock \bibinfo{title}{{Radioheliography}}, in: \bibinfo{editor}{{Alder},
  B.}, \bibinfo{editor}{{Fernbach}, S.}, \bibinfo{editor}{{Rotenberg}, M.}
  (Eds.), \bibinfo{booktitle}{Methods in Computational Physics. Volume 14 -
  Radio astronomy}, pp. \bibinfo{pages}{1--53}.
\bibitem[{{Landau} \& {Ghigo}(1984)}]{Landau1984}
\bibinfo{author}{{Landau}, R.}, \bibinfo{author}{{Ghigo}, F.D.},
  \bibinfo{year}{1984}.
\newblock \bibinfo{title}{{Data Compression Techniques for Astronomy}}, in:
  \bibinfo{editor}{{Klinglesmith}, D.A.} (Ed.),
  \bibinfo{booktitle}{Astronomical Microdensitometry Conference}, p.
  \bibinfo{pages}{255}.
\bibitem[{Lazio(2013)}]{lazio-2011}
\bibinfo{author}{Lazio, J.}, \bibinfo{year}{2013}.
\newblock \bibinfo{title}{{T}he {S}quare {K}ilometre {A}rray {D}esign
  {R}eference {M}ission: {SKA} phase 1}.
\newblock \bibinfo{type}{Technical Report}
  \bibinfo{number}{SCI-020.010.020-DRM-002}. JPL/SKA.
\bibitem[{Lemire \& Boytsov(2014)}]{Lemire-2012}
\bibinfo{author}{Lemire, D.}, \bibinfo{author}{Boytsov, L.},
  \bibinfo{year}{2014}.
\newblock \bibinfo{title}{Decoding billions of integers per second through
  vectorization}.
\newblock \bibinfo{journal}{Software: Practice and Experience}
  \bibinfo{volume}{45}, \bibinfo{pages}{1--29}.
\bibitem[{Li(2003)}]{li-2003}
\bibinfo{author}{Li, J.}, \bibinfo{year}{2003}.
\newblock \bibinfo{title}{Image compression: The mathematics of jpeg 2000}.
\newblock \bibinfo{journal}{Modern Signal Processing} \bibinfo{volume}{46},
  \bibinfo{pages}{185--221}.
\bibitem[{{LZ4}(2011)}]{lz4}
\bibinfo{author}{{LZ4}}, \bibinfo{year}{2011}.
\newblock \bibinfo{title}{{RealTime Data Compression: LZ4 explained}}.
\bibitem[{{Miller} \& {Lynch}(1976)}]{Miller1976}
\bibinfo{author}{{Miller}, W.H.}, \bibinfo{author}{{Lynch}, T.J.},
  \bibinfo{year}{1976}.
\newblock \bibinfo{title}{{On-board image compression for the RAE lunar
  mission}}.
\newblock \bibinfo{journal}{IEEE Transactions on Aerospace Electronic Systems}
  \bibinfo{volume}{12}, \bibinfo{pages}{327--335}.
\bibitem[{{Mink} et~al.(2014){Mink}, {Mann}, {Hanisch}, {Rots}, {Seaman},
  {Jenness}, {Thomas} \& {O'Mullane}}]{Mink2014}
\bibinfo{author}{{Mink}, J.}, \bibinfo{author}{{Mann}, R.G.},
  \bibinfo{author}{{Hanisch}, R.}, \bibinfo{author}{{Rots}, A.},
  \bibinfo{author}{{Seaman}, R.}, et~al., \bibinfo{year}{2014}.
\newblock \bibinfo{title}{{The Past, Present and Future of Astronomical Data
  Formats}}.
\newblock \bibinfo{journal}{ArXiv e-prints} \eprint{1411.0996}.
\bibitem[{{Mosquera} \& {Kochanek}(2011)}]{Mosquera2011}
\bibinfo{author}{{Mosquera}, A.M.}, \bibinfo{author}{{Kochanek}, C.S.},
  \bibinfo{year}{2011}.
\newblock \bibinfo{title}{{The Microlensing Properties of a Sample of 87 Lensed
  Quasars}}.
\newblock \bibinfo{journal}{The Astrophysical Journal} \bibinfo{volume}{738},
  \bibinfo{pages}{96}.
\newblock \eprint{1104.2356}.
\bibitem[{{Natusch}(2014)}]{Natusch2014}
\bibinfo{author}{{Natusch}, T.}, \bibinfo{year}{2014}.
\newblock \bibinfo{title}{{Application of Lossless Data Compression Techniques
  to Radio Astronomy Data flows}}.
\newblock \bibinfo{journal}{ArXiv e-prints} \eprint{1405.5634}.
\bibitem[{{Oguri} \& {Marshall}(2010)}]{Oguri2010}
\bibinfo{author}{{Oguri}, M.}, \bibinfo{author}{{Marshall}, P.J.},
  \bibinfo{year}{2010}.
\newblock \bibinfo{title}{{Gravitationally lensed quasars and supernovae in
  future wide-field optical imaging surveys}}.
\newblock \bibinfo{journal}{Monthly Notices of the Royal Astronomical Society}
  \bibinfo{volume}{405}, \bibinfo{pages}{2579--2593}.
\newblock \eprint{1001.2037}.
\bibitem[{{Pence} et~al.(2011){Pence}, {Seaman} \& {White}}]{Pence2011}
\bibinfo{author}{{Pence}, W.}, \bibinfo{author}{{Seaman}, R.},
  \bibinfo{author}{{White}, R.L.}, \bibinfo{year}{2011}.
\newblock \bibinfo{title}{{A New Compression Method for FITS Tables}}, in:
  \bibinfo{editor}{{Evans}, I.N.}, \bibinfo{editor}{{Accomazzi}, A.},
  \bibinfo{editor}{{Mink}, D.J.}, \bibinfo{editor}{{Rots}, A.H.} (Eds.),
  \bibinfo{booktitle}{Astronomical Data Analysis Software and Systems XX}, p.
  \bibinfo{pages}{493}.
\bibitem[{{Pence} et~al.(2000){Pence}, {White}, {Greenfield} \&
  {Tody}}]{Pence2000}
\bibinfo{author}{{Pence}, W.}, \bibinfo{author}{{White}, R.L.},
  \bibinfo{author}{{Greenfield}, P.}, \bibinfo{author}{{Tody}, D.},
  \bibinfo{year}{2000}.
\newblock \bibinfo{title}{{A FITS Image Compression Proposal}}, in:
  \bibinfo{editor}{{Manset}, N.}, \bibinfo{editor}{{Veillet}, C.},
  \bibinfo{editor}{{Crabtree}, D.} (Eds.), \bibinfo{booktitle}{Astronomical
  Data Analysis Software and Systems IX}, p. \bibinfo{pages}{551}.
\bibitem[{{Pence} et~al.(2010){Pence}, {White} \& {Seaman}}]{Pence2010}
\bibinfo{author}{{Pence}, W.D.}, \bibinfo{author}{{White}, R.L.},
  \bibinfo{author}{{Seaman}, R.}, \bibinfo{year}{2010}.
\newblock \bibinfo{title}{{Optimal Compression of Floating-Point Astronomical
  Images Without Significant Loss of Information}}.
\newblock \bibinfo{journal}{Publications of the Astronomical Society of the
  Pacific} \bibinfo{volume}{122}, \bibinfo{pages}{1065--1076}.
\newblock \eprint{1007.1179}.
\bibitem[{{Peters} \& {Kitaeff}(2014)}]{Peters-2014}
\bibinfo{author}{{Peters}, S.M.}, \bibinfo{author}{{Kitaeff}, V.V.},
  \bibinfo{year}{2014}.
\newblock \bibinfo{title}{{The impact of JPEG2000 lossy compression on the
  scientific quality of radio astronomy imagery}}.
\newblock \bibinfo{journal}{Astronomy and Computing} \bibinfo{volume}{6},
  \bibinfo{pages}{41--51}.
\newblock \eprint{1401.7433}.
\bibitem[{{Poindexter} et~al.(2007){Poindexter}, {Morgan}, {Kochanek} \&
  {Falco}}]{Poindexter2007}
\bibinfo{author}{{Poindexter}, S.}, \bibinfo{author}{{Morgan}, N.},
  \bibinfo{author}{{Kochanek}, C.S.}, \bibinfo{author}{{Falco}, E.E.},
  \bibinfo{year}{2007}.
\newblock \bibinfo{title}{{Mid-IR Observations and a Revised Time Delay for the
  Gravitational Lens System Quasar HE 1104-1805}}.
\newblock \bibinfo{journal}{The Astrophysical Journal} \bibinfo{volume}{660},
  \bibinfo{pages}{146--151}.
\newblock \eprint{astro-ph/0612045}.
\bibitem[{{Press}(1992)}]{Press1992}
\bibinfo{author}{{Press}, W.H.}, \bibinfo{year}{1992}.
\newblock \bibinfo{title}{{Wavelet-Based Compression Software for FITS
  Images}}, in: \bibinfo{editor}{{Worrall}, D.M.},
  \bibinfo{editor}{{Biemesderfer}, C.}, \bibinfo{editor}{{Barnes}, J.} (Eds.),
  \bibinfo{booktitle}{Astronomical Data Analysis Software and Systems I},
  p.~\bibinfo{pages}{3}.
\bibitem[{{Price} et~al.(2014){Price}, {Barsdell} \& {Greenhill}}]{Price2014}
\bibinfo{author}{{Price}, D.C.}, \bibinfo{author}{{Barsdell}, B.R.},
  \bibinfo{author}{{Greenhill}, L.J.}, \bibinfo{year}{2014}.
\newblock \bibinfo{title}{{Is HDF5 a good format to replace UVFITS?}}
\newblock \bibinfo{journal}{ArXiv e-prints} \eprint{1411.0507}.
\bibitem[{Price-Whelan \& Hogg(2010)}]{Price-Whelan-2010}
\bibinfo{author}{Price-Whelan, A.M.}, \bibinfo{author}{Hogg, D.W.},
  \bibinfo{year}{2010}.
\newblock \bibinfo{title}{{What bandwidth do I need for my image?}}
\newblock \bibinfo{journal}{Pubs Astron Soc Pac} \bibinfo{volume}{Vol. 122}.
\newblock \eprint{arXiv:0910.2375v3}.
\bibitem[{{Quinn} et~al.(2015){Quinn}, {Axelrod}, {Bird}, {Dodson}, {Szalay} \&
  {Wicenec}}]{Quinn2015}
\bibinfo{author}{{Quinn}, P.}, \bibinfo{author}{{Axelrod}, T.},
  \bibinfo{author}{{Bird}, I.}, \bibinfo{author}{{Dodson}, R.},
  \bibinfo{author}{{Szalay}, A.}, et~al., \bibinfo{year}{2015}.
\newblock \bibinfo{title}{{Delivering SKA Science}}.
\newblock \bibinfo{journal}{ArXiv e-prints} \eprint{1501.05367}.
\bibitem[{Salomon(2007)}]{DCC}
\bibinfo{author}{Salomon, D.}, \bibinfo{year}{2007}.
\newblock \bibinfo{title}{{Data Compression: The Complete Reference}}.
\newblock \bibinfo{publisher}{Springer-Verlag New York, Inc.},
  \bibinfo{address}{Secaucus, NJ, USA}.
\newblock \bibinfo{note}{With contributions by Giovanni Motta and David
  Bryant.}
\bibitem[{{Schneider} \& {Weiss}(1986)}]{Schneider1986}
\bibinfo{author}{{Schneider}, P.}, \bibinfo{author}{{Weiss}, A.},
  \bibinfo{year}{1986}.
\newblock \bibinfo{title}{{The two-point-mass lens - Detailed investigation of
  a special asymmetric gravitational lens}}.
\newblock \bibinfo{journal}{Astronomy and Astrophysics} \bibinfo{volume}{164},
  \bibinfo{pages}{237--259}.
\bibitem[{{Schneider} \& {Weiss}(1987)}]{Schneider1987}
\bibinfo{author}{{Schneider}, P.}, \bibinfo{author}{{Weiss}, A.},
  \bibinfo{year}{1987}.
\newblock \bibinfo{title}{{A gravitational lens origin for AGN-variability?
  Consequences of micro-lensing}}.
\newblock \bibinfo{journal}{Astronomy and Astrophysics} \bibinfo{volume}{171},
  \bibinfo{pages}{49--65}.
\bibitem[{{Seaman}(2011)}]{Seaman2011}
\bibinfo{author}{{Seaman}, R.}, \bibinfo{year}{2011}.
\newblock \bibinfo{title}{{Tile-Compressed FITS Kernel for IRAF}}, in:
  \bibinfo{editor}{{Evans}, I.N.}, \bibinfo{editor}{{Accomazzi}, A.},
  \bibinfo{editor}{{Mink}, D.J.}, \bibinfo{editor}{{Rots}, A.H.} (Eds.),
  \bibinfo{booktitle}{Astronomical Data Analysis Software and Systems XX}, p.
  \bibinfo{pages}{501}.
\bibitem[{Shamir \& Nemiroff(2005)}]{Lior2005}
\bibinfo{author}{Shamir, L.}, \bibinfo{author}{Nemiroff, R.J.},
  \bibinfo{year}{2005}.
\newblock \bibinfo{title}{{PHOTZIP : A Lossy FITS Image Compression Algorithm
  that Protects User-Defined Levels of Photometric Integrity serves Bright
  Signals}}.
\newblock \bibinfo{journal}{The Astronomical Journal} \bibinfo{volume}{Vol.
  129}, \bibinfo{pages}{539--546}.
\bibitem[{Shannon(1948)}]{shannon}
\bibinfo{author}{Shannon, C.}, \bibinfo{year}{1948}.
\newblock \bibinfo{title}{{A Mathematical Theory of Communication}}.
\newblock \bibinfo{journal}{Bell System Technical Journal}
  \bibinfo{volume}{Vol. 27}, \bibinfo{pages}{379--423}.
\bibitem[{{Springel} et~al.(2005){Springel}, {White}, {Jenkins}, {Frenk},
  {Yoshida}, {Gao}, {Navarro}, {Thacker}, {Croton}, {Helly}, {Peacock}, {Cole},
  {Thomas}, {Couchman}, {Evrard}, {Colberg} \& {Pearce}}]{Springel-2005}
\bibinfo{author}{{Springel}, V.}, \bibinfo{author}{{White}, S.D.M.},
  \bibinfo{author}{{Jenkins}, A.}, \bibinfo{author}{{Frenk}, C.S.},
  \bibinfo{author}{{Yoshida}, N.}, et~al., \bibinfo{year}{2005}.
\newblock \bibinfo{title}{{Simulations of the formation, evolution and
  clustering of galaxies and quasars}}.
\newblock \bibinfo{journal}{Nature} \bibinfo{volume}{435},
  \bibinfo{pages}{629--636}.
\newblock \eprint{astro-ph/0504097}.
\bibitem[{{Starck} et~al.(1995){Starck}, {Murtagh} \& {Louys}}]{Starck1995}
\bibinfo{author}{{Starck}, J.L.}, \bibinfo{author}{{Murtagh}, F.},
  \bibinfo{author}{{Louys}, M.}, \bibinfo{year}{1995}.
\newblock \bibinfo{title}{{Astronomical Image Compression Using the Pyramidal
  Median Transform}}, in: \bibinfo{editor}{{Shaw}, R.A.},
  \bibinfo{editor}{{Payne}, H.E.}, \bibinfo{editor}{{Hayes}, J.J.E.} (Eds.),
  \bibinfo{booktitle}{Astronomical Data Analysis Software and Systems IV}, p.
  \bibinfo{pages}{268}.
\bibitem[{Taubman(2005)}]{taubman-2005}
\bibinfo{author}{Taubman, D.}, \bibinfo{year}{2005}.
\newblock \bibinfo{title}{{Kakadu Survey Documentation} (last updated for
  version 5.0)}.
\bibitem[{{Thompson} et~al.(2010){Thompson}, {Fluke}, {Barnes} \&
  {Barsdell}}]{Thompson-2010}
\bibinfo{author}{{Thompson}, A.C.}, \bibinfo{author}{{Fluke}, C.J.},
  \bibinfo{author}{{Barnes}, D.G.}, \bibinfo{author}{{Barsdell}, B.R.},
  \bibinfo{year}{2010}.
\newblock \bibinfo{title}{{Teraflop per second gravitational lensing
  ray-shooting using graphics processing units}}.
\newblock \bibinfo{journal}{New Astronomy} \bibinfo{volume}{15},
  \bibinfo{pages}{16--23}.
\newblock \eprint{0905.2453}.
\bibitem[{{Vasilyev}(1998)}]{Vasilyev1998}
\bibinfo{author}{{Vasilyev}, S.V.}, \bibinfo{year}{1998}.
\newblock \bibinfo{title}{{An Optimal Data Loss Compression Technique for
  Remote Surface Multiwavelength Mapping}}, in: \bibinfo{editor}{{Albrecht},
  R.}, \bibinfo{editor}{{Hook}, R.N.}, \bibinfo{editor}{{Bushouse}, H.A.}
  (Eds.), \bibinfo{booktitle}{Astronomical Data Analysis Software and Systems
  VII}, p. \bibinfo{pages}{504}.
\bibitem[{{Veran} \& {Wright}(1994)}]{Veran1994}
\bibinfo{author}{{Veran}, J.P.}, \bibinfo{author}{{Wright}, J.},
  \bibinfo{year}{1994}.
\newblock \bibinfo{title}{{COMPRESSION Software for Astronomical Images}}, in:
  \bibinfo{editor}{{Crabtree}, D.R.}, \bibinfo{editor}{{Hanisch}, R.J.},
  \bibinfo{editor}{{Barnes}, J.} (Eds.), \bibinfo{booktitle}{Astronomical Data
  Analysis Software and Systems III}, p. \bibinfo{pages}{519}.
\bibitem[{{Vernardos} \& {Fluke}(2014)}]{Vernardos-2014}
\bibinfo{author}{{Vernardos}, G.}, \bibinfo{author}{{Fluke}, C.J.},
  \bibinfo{year}{2014}.
\newblock \bibinfo{title}{{Adventures in the microlensing cloud: Large
  datasets, eResearch tools, and GPUs}}.
\newblock \bibinfo{journal}{Astronomy and Computing} \bibinfo{volume}{6},
  \bibinfo{pages}{1--18}.
\newblock \eprint{1406.0559}.
\bibitem[{{Vernardos} et~al.(2014){Vernardos}, {Fluke}, {Bate} \&
  {Croton}}]{VerdanosEtAl2014}
\bibinfo{author}{{Vernardos}, G.}, \bibinfo{author}{{Fluke}, C.J.},
  \bibinfo{author}{{Bate}, N.F.}, \bibinfo{author}{{Croton}, D.},
  \bibinfo{year}{2014}.
\newblock \bibinfo{title}{{GERLUMPH Data Release 1: High-resolution
  Cosmological Microlensing Magnification Maps and eResearch Tools}}.
\newblock \bibinfo{journal}{The Astrophysical Journal Supplement}
  \bibinfo{volume}{211}, \bibinfo{pages}{16}.
\newblock \eprint{1401.7711}.
\bibitem[{Vohl(2013)}]{Vohl:Thesis:2013}
\bibinfo{author}{Vohl, D.}, \bibinfo{year}{2013}.
\newblock \bibinfo{title}{{Algorithmes de compression d'images hyperspectrales
  astrophysiques}}.
\newblock Master's thesis. Universit\'e Laval. \bibinfo{address}{Canada}.
\bibitem[{{White} \& {Percival}(1994)}]{white-1994}
\bibinfo{author}{{White}, R.L.}, \bibinfo{author}{{Percival}, J.W.},
  \bibinfo{year}{1994}.
\newblock \bibinfo{title}{{Compression and progressive transmission of
  astronomical images}}, in: \bibinfo{editor}{{Stepp}, L.M.} (Ed.),
  \bibinfo{booktitle}{Advanced Technology Optical Telescopes V}, pp.
  \bibinfo{pages}{703--713}.
\bibitem[{Ziv \& Lempel(1977)}]{LZ77}
\bibinfo{author}{Ziv, J.}, \bibinfo{author}{Lempel, A.}, \bibinfo{year}{1977}.
\newblock \bibinfo{title}{A universal algorithm for sequential data
  compression}.
\newblock \bibinfo{journal}{IEEE TRANSACTIONS ON INFORMATION THEORY}
  \bibinfo{volume}{23}, \bibinfo{pages}{337--343}.

\end{thebibliography}







\end{document}